\documentclass[lettersize,journal]{IEEEtran}
\usepackage{epsfig} 
\IEEEoverridecommandlockouts                              

\usepackage{graphicx} 
\usepackage{epsfig} 
\usepackage{amsmath} 
\usepackage{amssymb}  
\usepackage{subfig}
\usepackage[ruled,longend]{algorithm2e}
\usepackage{multicol}
\usepackage{color}
\usepackage[colorlinks,linkcolor=blue]{hyperref}
\usepackage{threeparttable}
\usepackage[export]{adjustbox}
\usepackage{graphicx} 
\usepackage{subcaption}
\usepackage{amsthm}
\usepackage{import}

\newtheorem{remark}{Remark}

\usepackage{scalerel}

\newcommand\reallywidehat[1]{\arraycolsep=0pt\relax%
\begin{array}{c}
\stretchto{
  \scaleto{
    \scalerel*[\widthof{\ensuremath{#1}}]{\kern-.5pt\bigwedge\kern-.5pt}
    {\rule[-\textheight/2]{1ex}{\textheight}} 
  }{\textheight} %
}{0.5ex}\\           
#1\\                 
\rule{-1ex}{0ex}
\end{array}
}

\begin{document}

\title{A Two-Stage, Model-Based Reinforcement Learning Approach for Active Flow Control of Bluff Body Wakes}



\author{Aayushman Sharma, Suman Chakravorty\thanks{Code available at: \url{https://github.com/AayushmanSharma96/Active-Flow-Control}}}


\maketitle
\begin{abstract}
This paper develops a data-driven, output-feedback approach to the infinite-horizon optimal control of high-dimensional nonlinear systems with unknown and unstable equilibria, using sparse partial observations. The approach builds on the transfer-plus-regulation decomposition of the infinite-horizon problem: a finite-horizon nonlinear transfer drives the system into a region where the dynamics are well-approximated by a linear model about the unknown operating point, and an infinite-horizon linear regulator identified within that region completes stabilization. We extend this framework to the partially observed setting by combining an ARMA-based information-state construction with a two-stage control architecture: an iterative linear quadratic regulator (iLQR) approach on the information state drives the system to the equilibrium neighborhood, discovered implicitly without prior knowledge of the target, and a locally identified time-invariant ARMA model provides the infinite-horizon regulator for asymptotic stabilization. The method requires no adjoint solver, reduced-order model, or full-state access. We validate the approach on high-fidelity Navier-Stokes simulations of the cylinder wake at $\mathrm{Re}=100$ using only eight surface pressure sensors, an order of magnitude fewer than recent model-based RL methods. The controller achieves complete suppression of vortex-shedding-induced lift oscillations and a $44\%$ reduction in total drag relative to the uncontrolled baseline.
\end{abstract}
\section{Introduction}

Active flow control aims to suppress unsteady wake dynamics such as vortex shedding, oscillatory loads, and excess drag through closed-loop feedback, with applications spanning aerodynamics, turbomachinery, and energy systems. The circular cylinder wake provides a canonical benchmark for these objectives. Above $\mathrm{Re}\approx 47$, the steady wake loses stability through a Hopf bifurcation, producing periodic vortex shedding (the Bénard-von Kármán street) that increases drag and induces oscillatory lift forces~\cite{williamson1996vortex,noack2003hierarchy}.

Suppressing this shedding requires stabilizing the steady symmetric wake, which is an \emph{unstable equilibrium} that is never visited in open-loop operation. Its location in the $O(10^5)$-dimensional state space is not accessible without specialized continuation or steady-state solvers. In addition, only sparse measurements are available, typically a small number of surface pressure sensors. Taken together, this leads to the following problem: stabilize an unknown and unstable equilibrium using only partial observations and without access to a model. In this setting, the standard ``regulate to a known setpoint'' formulation is not applicable.

\begin{figure}[t]
\centering
\subfloat[Uncontrolled wake at $\mathrm{Re}=100$: globally attracting vortex-shedding regime (limit cycle).]{
  \includegraphics[width=\linewidth]{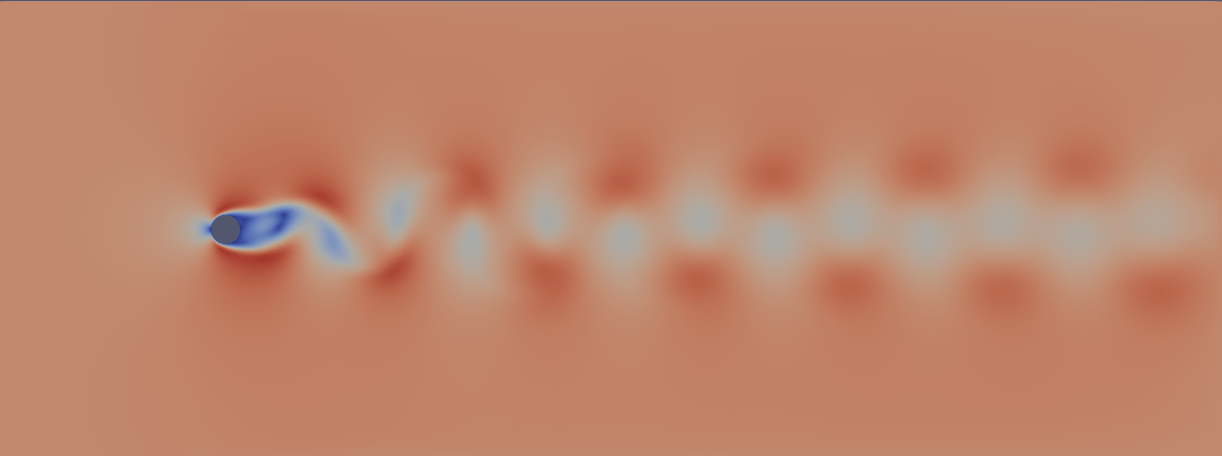}
  \label{fig:intro_uncontrolled}
}\\
\subfloat[Controlled wake: suppression of shedding and recovery of the steady symmetric wake (unstable equilibrium).]{
  \includegraphics[width=\linewidth]{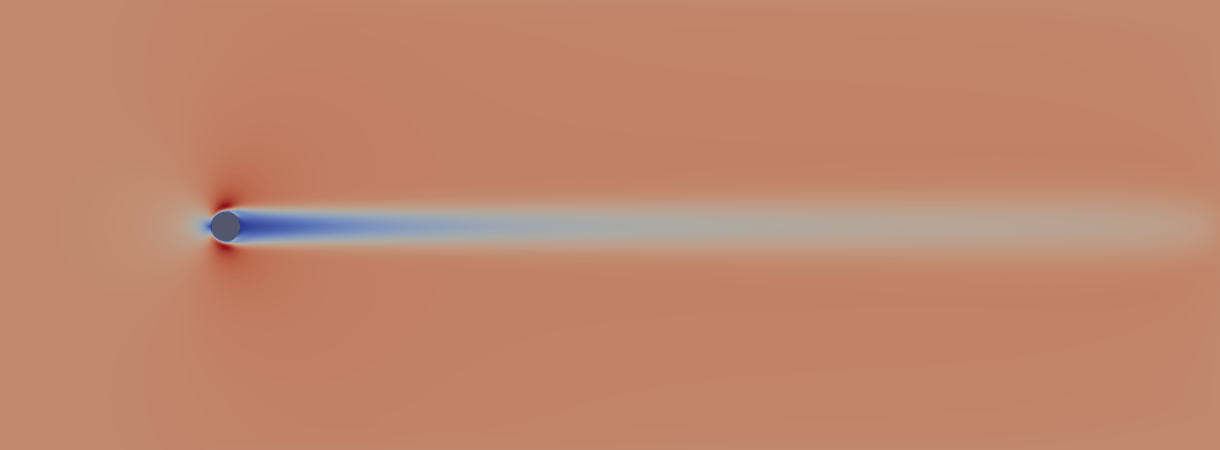}
  \label{fig:intro_controlled}
}
\caption{For $\mathrm{Re}>47$, the desired steady wake is unstable and not observed in open loop, while the dynamics converge to a vortex-shedding limit cycle.}
\label{fig:intro_motivation}
\end{figure}

Classical approaches construct reduced-order models, such as using POD or balanced truncation, and design controllers on these reduced order systems (\cite{rowley2005model,willcox2002balanced}), enabling wake
stabilization in a variety of
settings (\cite{bewley2001flow,Illingworth2016,Graham1999}). However, these methods require either an adjoint solver or access to the
full-state snapshots for model construction, and the resulting
controllers are tied to the fidelity of the reduced basis. In parallel,
deep reinforcement learning (DRL) has shown that vortex shedding can be suppressed directly from interaction data~\cite{rabault2019artificial}, with extensions to robustness and
related bluff-body geometries (\cite{tang2020robust,park2020deep,wang2024robust}).
However, model-free RL is sample-inefficient, which is a major
limitation when each interaction requires a high-fidelity CFD solve.
Model-based RL improves sample efficiency by training neural surrogates of the dynamics (\cite{ye2025mbrl,dong2023surrogate}), but these approaches rely on high-capacity models, dense sensor arrays (up to 151 probes~\cite{ye2025mbrl}), and do not provide closed-form guarantees on the resulting feedback law.

Across both classical and learning-based approaches, a fundamental
difficulty remains unresolved. The controller must both \emph{find} the
equilibrium and \emph{stabilize} it, using only partial observations
and without an explicit model. Most existing methods address one of
these aspects but not both simultaneously. \textit{In this regard, the simple yet fundamental insight we bring is that the infinite-horizon control problem admits a natural
decomposition into two tractable
subproblems (\cite{mohamed2023optimal,mohamed2025optimal}): a finite-horizon
nonlinear transfer that drives the system into a region where the
dynamics are well-approximated by a linear model about the unknown
operating point, and an infinite-horizon linear regulator that completes
stabilization within that region. The present work extends this framework to the cylinder wake flow setting, where (i)~the target equilibrium is unknown, (ii)~only sparse output measurements are available, and (iii)~both the transfer and the local model must be identified entirely from data}. To implement this idea, we propose a fully data-driven,
output-feedback framework that combines the information-state
construction of
POD2C (\cite{TNNLS_PODiLQR,wang2021data}) with the
transfer-plus-regulation architecture. In Stage~I, an iLQR
procedure operating on an ARMA-based information state performs
finite-horizon trajectory optimization, driving the flow from the
vortex-shedding limit cycle into the equilibrium neighborhood. The
equilibrium is not specified in advance but is discovered implicitly by
the optimizer. In Stage~II, a local time-invariant ARMA model
identified from data is used to design an infinite-horizon LQR
regulator for asymptotic stabilization. The method requires no adjoint
solver, reduced-order model, or full-state access.

We demonstrate the approach on a high-order PyFR simulation~\cite{witherden2014pyfr} of the cylinder wake at $\mathrm{Re}=100$ ($n_x \approx 10^5$) using only eight surface pressure sensors. The controller achieves complete suppression of lift oscillations and a $44\%$ reduction in total drag (pressure + viscous, full surface integration) relative to the uncontrolled baseline. This is achieved with significantly fewer sensors than recent RL-based approaches, while retaining a Riccati-based feedback structure. 

The rest of the paper is organized as follows. Section~\ref{sec2} presents the governing equations and nonlinear optimal control formulation. Section~\ref{sec3} discusses the unknown-equilibrium and partial-observation challenges and introduces the output construction. Section~\ref{sec4} presents the information-state modeling and the two-stage POD-iLQR/ARMA-LQR design. Section~\ref{sec5} presents numerical experiments.

\section{Problem Formulation}
\label{sec2}

\subsection{Cylinder Wake Flow Dynamics}
\label{subsec:dynamics}

We consider the two-dimensional incompressible cylinder wake flow governed by the
Navier-Stokes (NS) equations,
\begin{align}
    \frac{\partial \mathbf{v}}{\partial t}
        + (\mathbf{v} \cdot \nabla)\mathbf{v}
        &= -\nabla p + \frac{1}{\mathrm{Re}}\nabla^{2}\mathbf{v}
           + \mathbf{b}(\mathbf{u}), \label{eq:ns_mom} \\
    \nabla \cdot \mathbf{v} &= 0, \label{eq:ns_cont}
\end{align}
where $\mathbf{v}(\mathbf{x},t) \in \mathbb{R}^{2}$ is the velocity field,
$p(\mathbf{x},t)$ is the pressure field, $\mathrm{Re}$ is the Reynolds number,
and $\mathbf{b}(\mathbf{u})$ represents the body-force actuation parameterized
by the control input $\mathbf{u}(t)$.
At $\mathrm{Re} = 100$, the uncontrolled wake exhibits a supercritical Hopf
bifurcation, producing self-sustained periodic vortex shedding (the B\'{e}nard-von
K\'{a}rm\'{a}n street) about an \emph{unstable} steady-state equilibrium (Fig.~\ref{fig:intro_uncontrolled}).
\\

We discretize Eqs.~\eqref{eq:ns_mom}-\eqref{eq:ns_cont} in space and time using a
high-order spectral-element solver (PyFR \cite{witherden2014pyfr}), yielding a
finite-dimensional state $\mathbf{x}_{k} \in \mathbb{R}^{n_{x}}$ that
collects all pressure and velocities at the $k$-th
time step. The resulting discrete-time system takes the form
\begin{equation}
    \mathbf{x}_{k+1} = f(\mathbf{x}_{k}) + g(\mathbf{x}_{k})\,\mathbf{u}_{k},
    \label{eq:state_space}
\end{equation}
where $f:\mathbb{R}^{n_{x}}\!\rightarrow\!\mathbb{R}^{n_{x}}$ encodes the
nonlinear discretized NS dynamics and
$g:\mathbb{R}^{n_{x}}\!\rightarrow\!\mathbb{R}^{n_{x}\times n_{u}}$ is the
control influence matrix. Actuation is provided by two antisymmetric body-force inputs, 
parameterized by a single scalar $u_k \in \mathbb{R}$, with jet amplitudes $[u_k,\,-u_k]$. The precise actuator geometry and sensor placement are described in Section~\ref{sec5}, and given in Fig.~\ref{fig:setup}. 

\begin{remark}
    Note that the state dimension $n_{x}$ is typically on the order of $\mathcal{O}(10^{5}-10^{6})$, rendering full-state feedback infeasible in practice.
\end{remark}
\begin{remark}
    The system is assumed to be sufficiently discretized to accurately represent the dynamics of the Navier-Stokes equations. All optimality claims pertain to this high-dimensional yet finite representation.
\end{remark}

\subsection{Partial Observation Model}
\label{subsec:observation}
\begin{figure*}[t]
    \centering
    \includegraphics[width=0.7\textwidth]{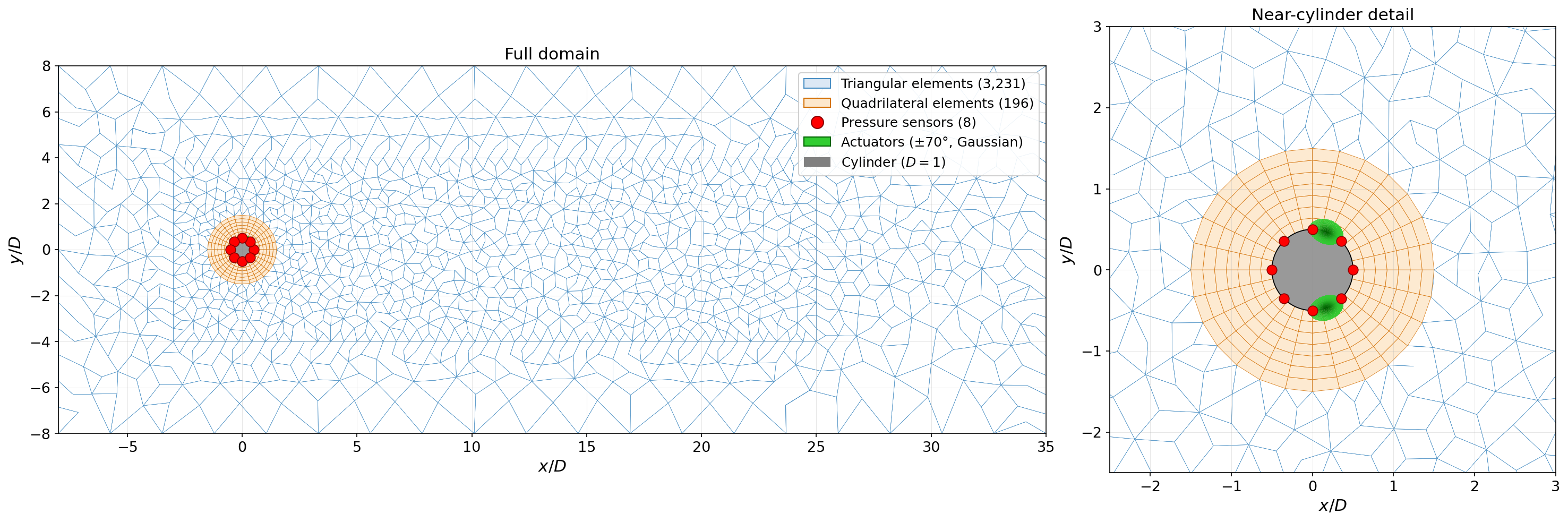}
    \caption{Computational mesh and control configuration.
    \textit{Left:} Full domain ($x \in [-8,35]D$, $y \in [-8,8]D$)
    with triangular (blue) and quadrilateral (orange) spectral elements
    ($p=3$, $3{,}427$ total). Red markers: surface pressure sensors;
    green: Gaussian actuator intensity at $\pm 70^\circ$.
    \textit{Right:} Near-cylinder detail showing the structured quad
    annulus, actuator profiles, and sensor placement.}
    \label{fig:setup}
\end{figure*}

In a practical flow-control setting, the full velocity/pressure field
$\mathbf{x}_{k}$ is not accessible. Instead, we assume that only
$n_{p} \ll n_{x}$ surface pressure measurements are available, obtained
from sensors placed on the cylinder surface. The pressure-sensor measurement model is
    $\mathbf{p}_{k} = h(\mathbf{x}_{k}),
    \label{eq:obs}$
where $\mathbf{p}_{k} \in \mathbb{R}^{n_{p}}$ and $h:\mathbb{R}^{n_{x}}
\rightarrow \mathbb{R}^{n_{p}}$ is a linear sampling operator that extracts
the $n_{p}$ surface pressure values from the state. Here, we consider $n_{p} = 8$ surface pressure sensors, spaced equally on the surface of the cylinder body (Fig.~\ref{fig:setup}). 

For feedback design, we subsequently map the raw pressure measurements $\mathbf{p}_k$ to the two-dimensional force-surrogate observation
$z_k \triangleq [\widehat{C}_D(\mathbf{p}_k),\,\widehat{C}_L(\mathbf{p}_k)]^\top \in \mathbb{R}^2$, which serves as the measured signal for the information-state construction and output-feedback control. The construction of $\widehat{C}_D(\cdot)$ and $\widehat{C}_L(\cdot)$ from the sparse pressure sensors is given in Section~\ref{subsec:force_est}.


Our overall objective is to find the infinite-horizon output-feedback control policy $\pi$ that depends only on the available observation history such that the closed-loop system is stabilized to the (unknown) operating point while minimizing drag, i.e., 
\begin{equation}
J^*_\infty(\mathbf{z}_0) = \min_{\pi}\; \sum_{k=0}^{\infty} c(\mathbf{z}_k, u_k),
\qquad u_k = \pi\!\bigl(\{z_i, u_i\}_{i=0}^{k-1}, z_k\bigr),
\label{eq:ihcost}
\end{equation}
subject to the flow dynamics (Eqs.~\eqref{eq:ns_mom}-\eqref{eq:ns_cont}) and the measurement model discussed above. Rather than attempting to solve Eq.~\eqref{eq:ihcost} directly, we decompose it into two stages: a finite-horizon nonlinear transfer that drives the system into a neighborhood of the equilibrium, and an infinite-horizon linear regulator that completes stabilization within that neighborhood. This decomposition is developed in Sections~\ref{sec3} and \ref{sec4}, following the transfer-plus-regulation architecture of~\cite{mohamed2023optimal}, adapted here to the output-feedback, unknown-equilibrium setting.

\section{Challenges in Data-Driven Flow Control}
\label{sec3}

\subsection{The Unknown Target Equilibrium}
\label{subsec:unknown_equil}

A fundamental difficulty that distinguishes flow control from standard
nonlinear control/ robotic control problems is that the desired target equilibrium
is not known \textit{a priori}. At $\mathrm{Re} > 47$, the steady symmetric wake
(the state of zero vortex shedding) is an \emph{unstable} fixed point of Eq.~\eqref{eq:state_space} that is never visited in open-loop operation. Its
exact location in the high-dimensional state space $\mathbb{R}^{n_{x}}$ is
therefore inaccessible from simulation or experiment without special
continuation methods. This renders the standard tracking formulation (which requires an explicit target state $\mathbf{x}^{*}$ in the cost)
infeasible.

A classical workaround is to define the cost in terms of physically meaningful scalar quantities that are zero at the desired operating point.
The lift coefficient $C_{L}$ and drag coefficient $C_{D}$ serve precisely this role: at the unstable steady wake, lift vanishes ($C_{L} = 0$) and drag attains its minimum ($C_{D} \approx C_{D}^{\min}$). Driving both
quantities toward zero thus implicitly defines the target without requiring knowledge of the underlying state $\mathbf{x}^{*}$.

\subsection{Data-Driven Force Estimation and the Two-Stage Strategy}
\label{subsec:two_stage}

A further complication is that the lift and drag coefficients $(C_L,C_D)$ depend on the full surface pressure and shear-stress distributions and therefore cannot be computed exactly from the sparse pressure measurements $\mathbf{p}_k$. To obtain physically meaningful outputs for feedback, we construct data-driven approximations $\widehat{C}_D(\mathbf{p}_k)$ and $\widehat{C}_L(\mathbf{p}_k)$ of the \emph{pressure} contributions to drag and lift from open-loop simulation data (Section~\ref{subsec:force_est}). These surrogates provide an interpretable coordinate system in which the objective becomes explicit: drive $\widehat{C}_L$ to zero (the equilibrium condition) while reducing $\widehat{C}_D$.

Sparse sensing introduces an additional challenge: the instantaneous measurement $\mathbf{z}_k \triangleq \bigl[\widehat{C}_D(\mathbf{p}_k),\,\widehat{C}_L(\mathbf{p}_k)\bigr]^\top$ is generally not Markov, so controllers based on local dynamic programming (e.g., iLQR/LQR) cannot be applied directly in measurement space. We therefore follow an ARMA-based \emph{information-state} construction and define an augmented state $Z_k$ by stacking a short history of outputs and inputs. Under mild observability conditions, this augmentation yields an approximately Markov representation that is sufficient for identifying local input-output linear models directly from rollouts and for performing Riccati-based backward recursions \cite{TNNLS_PODiLQR}. 

Even with this representation, the unknown target equilibrium may lie far from the initial vortex-shedding regime, and the equilibrium itself is unstable. This motivates the two-stage strategy adopted in this paper:



\paragraph{\textbf{Stage~I (global transfer via POD-iLQR)}}
We apply an information-state-based iterative LQR algorithm to perform
finite-horizon trajectory optimization starting from the shedding limit
cycle. Because the optimization searches over full nonlinear
trajectories rather than relying on a single local linearization, it
can navigate the nonconvex landscape and drive the flow into a
neighborhood of the unknown equilibrium without explicit knowledge of
its location.

\paragraph{ \textbf{Stage~II (local stabilization via ARMA-LQR)}}
Once the Stage~I trajectory reaches the equilibrium neighborhood, the
dynamics are well-approximated locally by a time-invariant ARMA model
identified from input-output rollouts. We then design an
infinite-horizon LQR regulator for this local model, which provides
asymptotic stabilization about the recovered operating point.

\noindent The formal development is given in Section~\ref{sec4}.

\begin{remark}
    The two-stage strategy described in Section~\ref{sec3} is an
instantiation of the \emph{transfer-plus-regulation} framework for
infinite-horizon nonlinear optimal control developed
in~\cite{mohamed2023optimal, mohamed2025optimal}. Under full-state access, linear controllability of the equilibrium, and nonlinear reachability of a terminal set~\cite{mohamed2023optimal}, that framework decomposes the
infinite-horizon problem into a finite-horizon nonlinear transfer that
drives the state into a terminal set, followed by an LQR regulator that
completes stabilization within the set; the composite cost-to-go is
shown to be a Control Lyapunov Function (CLF) and to converge to the
true infinite-horizon optimum as the terminal set
shrinks~\cite{mohamed2023optimal}. The present work extends this
framework to the partially observed setting, where (i)~the target
equilibrium is \emph{unknown}, (ii)~only sparse output measurements are
available, and (iii)~both the transfer and the local model must be identified from data.
\end{remark}

\section{Information-State based Data-driven Flow Control Approach}
\label{sec4}
\subsection{Force Coefficient Estimation from Sparse Pressure Sensors}
\label{subsec:force_est}
The lift and drag coefficients $C_L$ and $C_D$ depend on the full surface pressure and shear-stress distribution, neither of which is directly accessible from the $n_p = 8$ surface pressure sensors. We approximate the \emph{pressure contributions} to these coefficients via trapezoidal quadrature over the sensor locations. Let $\{\theta_i\}_{i=1}^{n_p}$ denote the equally-spaced azimuthal positions of the sensors, with $\Delta\theta = 2\pi/n_p$. Then
\begin{align}
    \widehat{C}_D(\mathbf{p}_k)
        &= -\frac{R\,\Delta\theta}{q_\infty D}
           \sum_{i=1}^{n_p} p_i^{(k)}\cos\theta_i,
    \label{eq:cdhat} \\
    \widehat{C}_L(\mathbf{p}_k)
        &= -\frac{R\,\Delta\theta}{q_\infty D}
           \sum_{i=1}^{n_p} p_i^{(k)}\sin\theta_i,
    \label{eq:clhat}
\end{align}
where $p_i^{(k)}$ is the pressure at sensor $i$ at time step $k$, $R = D/2$ is the cylinder radius, $D = 1$ is the cylinder diameter, and $q_\infty = \tfrac{1}{2}\rho U_\infty^2$ is the free-stream dynamic pressure. These approximations omit the viscous (skin-friction)
contribution to drag; at $\mathrm{Re} = 100$, the pressure drag dominates the total drag, so Eqs.~\eqref{eq:cdhat}-\eqref{eq:clhat} provide
a physically consistent, computationally inexpensive surrogate.
In what follows, we define the two-dimensional observation used for control as
\begin{equation}
    \mathbf{z}_k \triangleq \bigl[\widehat{C}_D(\mathbf{p}_k),\,\widehat{C}_L(\mathbf{p}_k)\bigr]^\top \in \mathbb{R}^{n_z},
    \qquad n_z = 2,
\end{equation}
where $\mathbf{p}_k \in \mathbb{R}^{n_p}$ denotes the raw pressure-sensor vector.

\subsection{Information-State Formulation}
\label{subsec:info_state}

Since the full flow state $\mathbf{x}_k$ is inaccessible, we follow the ARMA-based
information-state construction of~\cite{TNNLS_PODiLQR}. Under mild observability conditions, the output deviation $\delta\mathbf{z}_k = \mathbf{z}_k - \bar{\mathbf{z}}_k$
from a nominal trajectory can be expressed as a linear function of a
finite history of past output and control deviations through a time-varying ARMA model of order~$q$~\cite{TNNLS_PODiLQR}:
\begin{align}
    \nonumber \delta z_{k} = &\alpha_{k-1}  \delta z_{k-1}+\cdots+\alpha_{k-q}   \delta z_{k-q}   + \\& \beta_{k-1} \delta u_{k-1}+\cdots+ \beta_{k-q}   \delta u_{k-q},
    \label{eq:arma_io}
\end{align}
where $\delta u_k = u_k - \bar{u}_k$, and
$\alpha_{k-i} \in \mathbb{R}^{n_z \times n_z}$,
$\beta_{k-i} \in \mathbb{R}^{n_z \times n_u}$ are the ARMA coefficients
identified from data at each time step~$k$.
Defining the augmented information state
\begin{equation}
    \mathcal{Z}_k
        = \bigl[\mathbf{z}_k^\top,\,\mathbf{z}_{k-1}^\top,\,\ldots,\,
                \mathbf{z}_{k-q+1}^\top,\, u_{k-1},\,\ldots,\,
                 u_{k-q+1}\bigr]^\top
        \in \mathbb{R}^{n_\mathrm{aug}},
    \label{eq:info_state}
\end{equation}
with $n_\mathrm{aug} = n_z q + n_u(q - 1)$,
the ARMA model (Eq.~\eqref{eq:arma_io}) takes the companion form
\begin{equation}
    \delta\mathcal{Z}_{k} = \mathcal{A}_{k-1}\,\delta\mathcal{Z}_{k-1} + \mathcal{B}_{k-1}\,\delta u_{k-1},
    \label{eq:arma_dyn}
\end{equation}
The companion matrices $\mathcal{A}_k$ and $\mathcal{B}_k$ take the
structured form shown in Eq.~\eqref{eq:companion}.

\begin{table*}[!tp]
\begin{align}
\setcounter{MaxMatrixCols}{20}
\underbrace{
\begin{bmatrix}
\delta z_{k} \\ \delta z_{k-1} \\ \vdots \\ \delta z_{k-q+1} \\
\hline
\delta u_{k-1} \\ \vdots \\ \delta u_{k-q+1}
\end{bmatrix}}_{\delta\mathcal{Z}_k}
=
\underbrace{\begin{bmatrix}
\alpha_{k-1} & \alpha_{k-2} & \cdots & \alpha_{k-q+1} & \alpha_{k-q} 
& \vline &
\beta_{k-2} & \beta_{k-3} & \cdots & \beta_{k-q+1} & \beta_{k-q}\\
I & 0 & \cdots & 0 & 0 & \vline & 0 & 0 & \cdots & 0 & 0\\
0 & I & \cdots & 0 & 0 & \vline & 0 & 0 & \cdots & 0 & 0\\
\vdots & \ddots & \ddots & \vdots & \vdots & \vline & \vdots & \ddots & \ddots & \vdots & \vdots\\
0 & 0 & \cdots & I & 0 & \vline & 0 & 0 & \cdots & 0 & 0 \\
\hline
0 & 0 & \cdots & 0 & 0 & \vline & 0 & 0 & \cdots & 0 & 0\\
0 & 0 & \cdots & 0 & 0 & \vline & I & 0 & \cdots & 0 & 0\\
0 & 0 & \cdots & 0 & 0 & \vline & 0 & I & \cdots & 0 & 0\\
\vdots & & & \vdots & \vdots & \vline & \vdots & & \ddots & \vdots & \vdots \\
0 & 0 & \cdots & 0 & 0& \vline & 0 & 0 & \cdots & I & 0
\end{bmatrix}}_{\mathcal{A}_{k-1}}
\underbrace{
\begin{bmatrix}
\delta z_{k-1} \\ \vdots \\ \delta z_{k-q} \\
\hline
\delta u_{k-2} \\ \vdots \\ \delta u_{k-q}
\end{bmatrix}}_{\delta\mathcal{Z}_{k-1}}
+
\underbrace{
\begin{bmatrix}
\beta_{k-1} \\ 0 \\ \vdots \\ 0 \\
\hline
I \\ 0 \\ \vdots \\ 0
\end{bmatrix}}_{\mathcal{B}_{k-1}}
\delta u_{k-1}
\label{eq:companion}
\end{align}

\end{table*}

The top $n_z$ rows of $\mathcal{A}_{k-1}$ (coefficients $\alpha_{k-1},\ldots,\alpha_{k-q}$,
$\beta_{k-2},\ldots,\beta_{k-q}$) and the leading entry of $\mathcal{B}_{k-1}$
($\beta_{k-1}$) are identified from data; the remaining rows are fixed
companion-shift blocks that propagate the observation and control histories
forward in time.
\begin{remark}
    Although the discretized wake dynamics are extremely high-dimensional,
    cylinder wakes are well-known to evolve on low-dimensional manifolds
    dominated by a small number of coherent structures, as evidenced by POD-
    and balanced-POD-based reduced-order
    models~(\cite{noack2003hierarchy,rowley2005model,willcox2002balanced}).
    This structure suggests that only a modest amount of input-output history
    may be needed to form an approximately Markov information state for
    prediction and control. Consistent with this intuition, we use short
    histories ($q=3$) throughout
    (Tables~\ref{tab:params_stage1},~\ref{tab:params_stage2}).
\end{remark}

The first stage applies the POD-iLQR algorithm to compute the optimal open-loop control sequence by solving the finite-horizon optimal control problem 
\begin{equation}
   \min_{\{u_k\}_{k=0}^{N-1}}
    \sum_{k=0}^{N-1} c\bigl(\mathcal{Z}_k, u_k\bigr)
    + c_N\bigl(\mathcal{Z}_N\bigr),
    \label{eq:fhocp}
\end{equation}
with horizon $N = 50$ steps ($25$\,s at $\Delta t = 0.5$\,s).
POD-iLQR is an iterative procedure that alternates between a forward
simulation pass, an ARMA LTV system identification step, a backward
dynamic programming pass, and a control update.
The running cost takes the form
\begin{equation*}
   c(\mathcal{Z}_k, u_k) = \mathcal{Z}_k^\top Q\,\mathcal{Z}_k + u_k^\top R\,u_k,   
\end{equation*}
where $Q$ is block-diagonal over the observation history with a
weighting that penalizes $\widehat{C}_L$ more heavily than $\widehat{C}_D$,
reflecting that $\widehat{C}_L = 0$ is a hard equilibrium condition while
$\widehat{C}_D$ is treated as a soft performance objective. A scaled terminal
cost $c_N = \mathcal{Z}_N^\top Q_N\,\mathcal{Z}_N$ strongly incentivizes the trajectory to approach the unknown
equilibrium by the end of the horizon.

\subsubsection*{Step 1: Forward Pass}
Given the current nominal control sequence
$\bar{\mathbf{u}} = \{\bar{u}_k\}_{k=0}^{N-1}$, the PyFR simulator is
rolled out from the initial flow state to produce the nominal
information-state trajectory
$\bar{\mathcal{Z}} = \{\bar{\mathcal{Z}}_k\}_{k=0}^{N}$.

\subsubsection*{Step 2: ARMA LTV System Identification}
To identify the LTV matrices
$\{\mathcal{A}_k,\mathcal{B}_k\}_{k=0}^{N-1}$ in~\eqref{eq:arma_dyn},
we interact with the black-box purely through input-output
perturbations: $n_\mathrm{samp}$ rollouts are executed from the nominal
trajectory of Step~1, each with i.i.d.\ control perturbations
$\delta u_k^{(j)} \sim \mathcal{N}(0,\,\sigma^2)$ superimposed on
$\bar{u}_k$, and the resulting output deviations
$\delta\mathbf{z}_k^{(j)}$ are recorded. The collected output and control deviations are
assembled into the information-state deviations
\begin{equation*}
   \delta\mathcal{Z}_k = \mathcal{Z}_k - \bar{\mathcal{Z}}_k   
\end{equation*}
and fitted by SVD-truncated least squares, independently at each time step~$k$, to
obtain the companion matrices in~\eqref{eq:arma_dyn}.

\subsubsection*{Step 3: Backward Pass}
With the identified LTV model in hand, a backward Riccati recursion is
solved from $k = N$ to $k = 0$. Starting from the terminal conditions
$v_N = \frac{\partial c_N}{\partial \mathcal{Z}}\big|_{\mathcal{Z}_N}$
and $V_N = \nabla^2_{\mathcal{ZZ}} c_N\big|_{\mathcal{Z}_N}$, the
feedforward-feedback gain pair $(\kappa_k, K_k)$ is computed as
\begin{align}
    \kappa_k &= (R + \mathcal{B}_k^\top V_{k+1}\,\mathcal{B}_k)^{-1}
                (R\,\bar{u}_k + \mathcal{B}_k^\top v_{k+1}), \label{eq:kappa}\\
    K_k      &= (R + \mathcal{B}_k^\top V_{k+1}\,\mathcal{B}_k)^{-1}
                \mathcal{B}_k^\top V_{k+1}\,\mathcal{A}_k, \label{eq:Kgain}
\end{align}
with the value-function gradient and Hessian updated via
\begin{align}
    v_k = l_k^{\mathcal{Z}} + \mathcal{A}_k^\top v_{k+1}
          - \mathcal{A}_k^\top V_{k+1}\,\mathcal{B}_k
            (R + \mathcal{B}_k^\top V_{k+1}\,\mathcal{B}_k)^{-1} \nonumber\\
\cdot(\mathcal{B}_k^\top v_{k+1} + R\,\bar{u}_k),  
\label{eq:vk}\\[3pt]
    V_k = l_k^{\mathcal{ZZ}} + \mathcal{A}_k^\top V_{k+1}\,\mathcal{A}_k - \mathcal{A}_k^\top V_{k+1}\,\mathcal{B}_k (R + \mathcal{B}_k^\top V_{k+1}\,\mathcal{B}_k)^{-1} \nonumber\\
        \cdot\mathcal{B}_k^\top V_{k+1}\,\mathcal{A}_k, \label{eq:Vk}
\end{align}
where $l_k^{\mathcal{Z}} = \frac{\partial c}{\partial \mathcal{Z}}\big|_{(\bar{\mathcal{Z}}_k, \bar{u}_k)}$ and $l_k^{\mathcal{ZZ}} = \nabla^2_{\mathcal{ZZ}} c\big|_{(\bar{\mathcal{Z}}_k, \bar{u}_k)}$.
\\

\subsubsection*{Step 4: Control Update}
The nominal control sequence is updated as
\begin{equation}
    \bar{u}_k^{(i+1)} = \bar{u}_k^{(i)}
        + \alpha\,\kappa_k
        + K_k\bigl(\mathcal{Z}_k^{(i+1)} - \mathcal{Z}_k^{(i)}\bigr),
    \label{eq:ilqr_update}
\end{equation}
where $\alpha \in (0,1]$ is the line-search step size and
$\mathcal{Z}_0^{(i+1)} = \mathcal{Z}_0^{(i)}$. Steps~1--4 are repeated
until convergence.

\subsubsection*{Transfer Outcome}
At the end of the iLQR horizon, the flow has been driven from sustained periodic shedding to the neighborhood of the unknown equilibrium (Fig.~\ref{fig:vort_ilqr1}-\ref{fig:vort_ilqr3}).
The terminal drag value
$\widehat{C}_{D,N} \triangleq \widehat{C}_D(p_N)$ is recorded as the drag level achieved at the end of the finite-horizon transfer (Fig.~\ref{fig:results_cd}); it serves as an empirical indicator of the equilibrium neighborhood reached by Stage~I, rather than the final regulated drag level, which Stage~II drives lower. Crucially, $\widehat{C}_{D,N}$ is discovered implicitly by the nonlinear optimizer and does not require prior knowledge of the equilibrium state~$x^*$.

\subsection{Stage~II: ARMA-LQR Local Stabilization}
\label{subsec:stage2}
The second stage identifies an LTI model around the terminal operating
point and designs an LQR regulator to complete stabilization to the
unknown equilibrium. This stage implements the \emph{regulation}
component of the transfer-plus-regulation framework
in the data-driven, output-feedback setting.

\subsubsection*{LTI System Identification}
Starting from the terminal PyFR snapshot at time $t_N$, we execute
$n_r$ independent short rollouts of horizon $H$ steps under small random
control perturbations with standard deviation $\sigma$. The information-state
deviations from the terminal operating point are collected, and a single
time-invariant companion pair $(\mathcal{A}, \mathcal{B})$ is fitted by
SVD-truncated least squares across all transitions from all rollouts:
\begin{equation}
    \delta\mathcal{Z}_{k}
        = \mathcal{A}\,\delta\mathcal{Z}_{k-1} + \mathcal{B}\,\delta u_{k-1},
    \label{eq:lti_model}
\end{equation}
where $\delta\mathcal{Z}_k = \mathcal{Z}_k - \mathcal{Z}^*$ denotes the
deviation from the (zero) regulation reference $\mathcal{Z}^* = 0$, i.e.,
$z^* = [0,\,0]^\top$ in $(\widehat{C}_D,\widehat{C}_L)$ coordinates. 
All lagged observation and control slots are initialized to their corresponding
values at the terminal snapshot. The SVD truncation retains singular values that
explain 99.99\% of the total energy, suppressing poorly excited directions
without introducing bias.

\subsubsection*{LQR Design}
Given the identified model $(\mathcal{A},\mathcal{B})$, the
infinite-horizon LQR gain is computed by solving the
Discrete Algebraic Riccati Equation (DARE):
\begin{equation}
    K = \bigl(R_\ell + \mathcal{B}^\top P\,\mathcal{B}\bigr)^{-1}
                 \mathcal{B}^\top P\,\mathcal{A},
    \label{eq:lqr_gain}
\end{equation}
where $P$ satisfies the DARE for the pair $(\mathcal{A},\mathcal{B},Q_\ell,R_\ell)$.
The weight matrices $Q_\ell$ and $R_\ell$ follow the same block-diagonal
structure as Stage~I but with increased asymmetry toward $\widehat{C}_L$
and a relaxed control penalty, giving the regulator greater authority in
the local linear regime (see Table~\ref{tab:params_stage2}).
The resulting output feedback law is:
    $u_k = -K\,\bigl(\mathcal{Z}_k - \mathcal{Z}^*\bigr),
    \label{eq:lqr_law}$
which drives the information-state, and hence the force coefficients
$(\widehat{C}_D, \widehat{C}_L)$, toward the setpoint $(0, 0)$ in the quadratic objective.

\begin{remark}
    The regulation reference $\mathcal{Z}^*$ is fixed at zero, while the local ARMA model used in Stage~II is identified from rollouts initialized at the terminal Stage~I snapshot, i.e., within the equilibrium neighborhood reached by the transfer stage. This enables local stabilization without requiring knowledge of the full equilibrium state $x^*$.
\end{remark}

\section{Empirical Results}
\label{sec5}

\subsection{Experimental Setup}
\label{subsec:setup}

All simulations are performed using PyFR~\cite{witherden2014pyfr}, a
high-order spectral-element CFD solver based on the artificial
compressibility formulation of the incompressible Navier-Stokes
equations. The computational domain is a two-dimensional channel with a
circular cylinder of diameter $D = 1$ centered at the origin, with
freestream velocity $U_\infty = 1$ giving $\mathrm{Re} = 100$. The mesh consists of $3{,}427$ elements ($3{,}231$ triangular and $196$ quadrilateral) at solution polynomial order $p = 3$, yielding a state
dimension of $n_x = 106{,}338$ (35,446 solution points $\times$ 3 variables). Time
integration uses an SDIRK33 dual time-stepping scheme with physical timestep $\delta t = 0.1$\,s; the control timestep is $\Delta t = 0.5$\,s, corresponding to $5$ solver steps per control interval.

Actuation is implemented as two antisymmetric body-force inputs applied to the momentum equations in the cross-stream ($v$) direction, following the geometry of the reference~\cite{Illingworth2016}. Each actuator has a Gaussian profile centered at $\pm 70^\circ$ from the downstream horizontal with a $1\%$-cutoff half-width of $\pm 35^\circ$, and a
radial Gaussian with standard deviation $\sigma_r = 0.1R$ concentrated at the cylinder surface ($R = D/2$). The spatial distribution $G(\mathbf{x})$ is normalized so that
$\int G\,\mathrm{d}A = 1$, making the scalar amplitude $u_k \in \mathbb{R}$ equal to the net body-force magnitude per unit span integrated over the actuator footprint, with units of $U_\infty^2D$.
The two actuators are parameterized by a single scalar with antisymmetric amplitudes $[u_k,\,-u_k]$, corresponding to antisymmetric blowing/suction at $\pm 70^\circ$. Observations consist of $n_p = 8$ surface pressure sensors equally spaced on the cylinder surface, from which the force coefficient approximations $\widehat{C}_D$ and $\widehat{C}_L$ are
computed via Eqs.~\eqref{eq:cdhat}-\eqref{eq:clhat}. The mesh, actuator profiles, and sensor layout are illustrated in Fig.~\ref{fig:setup}.

The POD-iLQR and ARMA-LQR algorithm parameters for both stages are
reported in Tables~\ref{tab:params_stage1} and~\ref{tab:params_stage2}.
All experiments are initialized from a fully developed vortex shedding
state obtained by running the uncontrolled simulation to $t = 220$\,s,
well past transient onset.

\begin{table}[htbp]
\caption{POD-iLQR (Stage~I) algorithm parameters.}
\label{tab:params_stage1}
\centering
\begin{threeparttable}
\setlength{\tabcolsep}{1.2mm}{
\begin{tabular}{|c|c|}
\hline
\textbf{Parameter} & \textbf{Value} \\
\hline
Horizon $N$ (steps) & 50 \\
\hline
Control timestep $\Delta t$ (s) & 0.5 \\
\hline
History length $q$ & 3 \\
\hline
SysID rollouts $n_\mathrm{samp}$ & 30 \\
\hline
SysID perturbation $\sigma$ & 0.1 \\
\hline
State cost $Q$ & $\mathrm{diag}(50,500,50,500,50,500,0,0)$ \\
\hline
Terminal cost $Q_N$ & 20*Q \\
\hline
Control penalty $R$ & 100 \\
\hline
iLQR iterations & 7 \\
\hline
SVD energy threshold & 99.99\% \\
\hline
\end{tabular}
}
\end{threeparttable}
\end{table}

\begin{table}[htbp]
\caption{ARMA-LQR (Stage~II) algorithm parameters.}
\label{tab:params_stage2}
\centering
\begin{threeparttable}
\setlength{\tabcolsep}{1.2mm}{
\begin{tabular}{|c|c|}
\hline
\textbf{Parameter} & \textbf{Value} \\
\hline
LQR horizon $N_\ell$ (steps) & 200 \\
\hline
Control timestep $\Delta t$ (s) & 0.5 \\
\hline
History length $q$ & 3 \\
\hline
SysID rollouts $n_r$ & 10 \\
\hline
SysID rollout horizon $H$ (steps) & 50 \\
\hline
State cost $Q_\ell$ & $\mathrm{diag}(5,500,5,500,5,500,0,0)$ \\
\hline
Control penalty $R_\ell$ & 5 \\
\hline
\end{tabular}
}
\end{threeparttable}
\end{table}

\subsection{Uncontrolled Flow Baseline}
\label{subsec:baseline}

\begin{figure}[t]
    \centering

    \subfloat[Instantaneous streamwise velocity field $u$ showing the
    Bénard-von Kármán street.]
    {
        \parbox{0.9\linewidth}{
            \centering
            \includegraphics[width=0.7\linewidth]{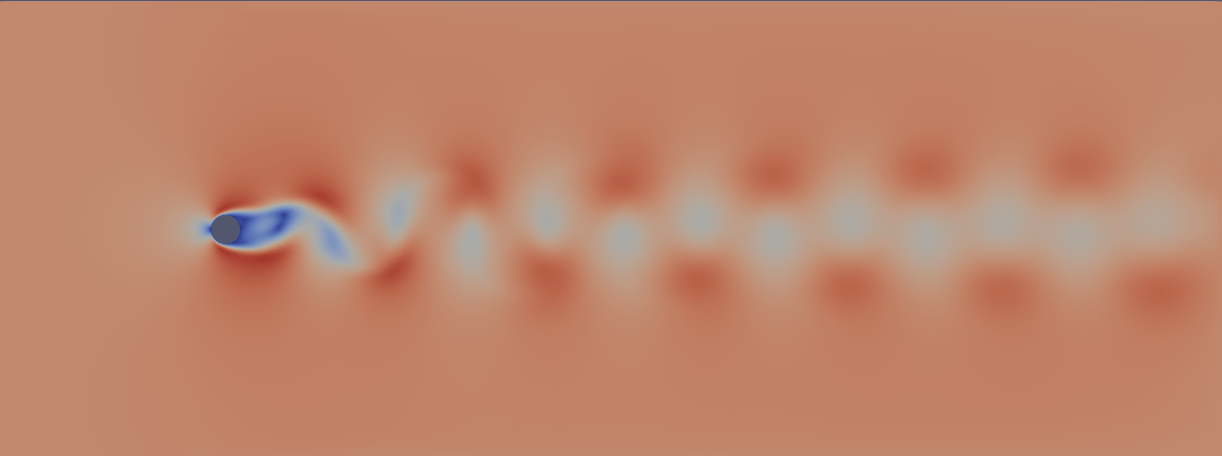}
            \label{fig:baseline_vorticity}
        }
    }


    \subfloat[Time series of $\widehat{C}_D$ (blue) and $\widehat{C}_L$
    (red) on the pressure-approximated limit cycle.]
    {
        \parbox{0.9\linewidth}{
            \centering
            \includegraphics[width=0.9\linewidth]{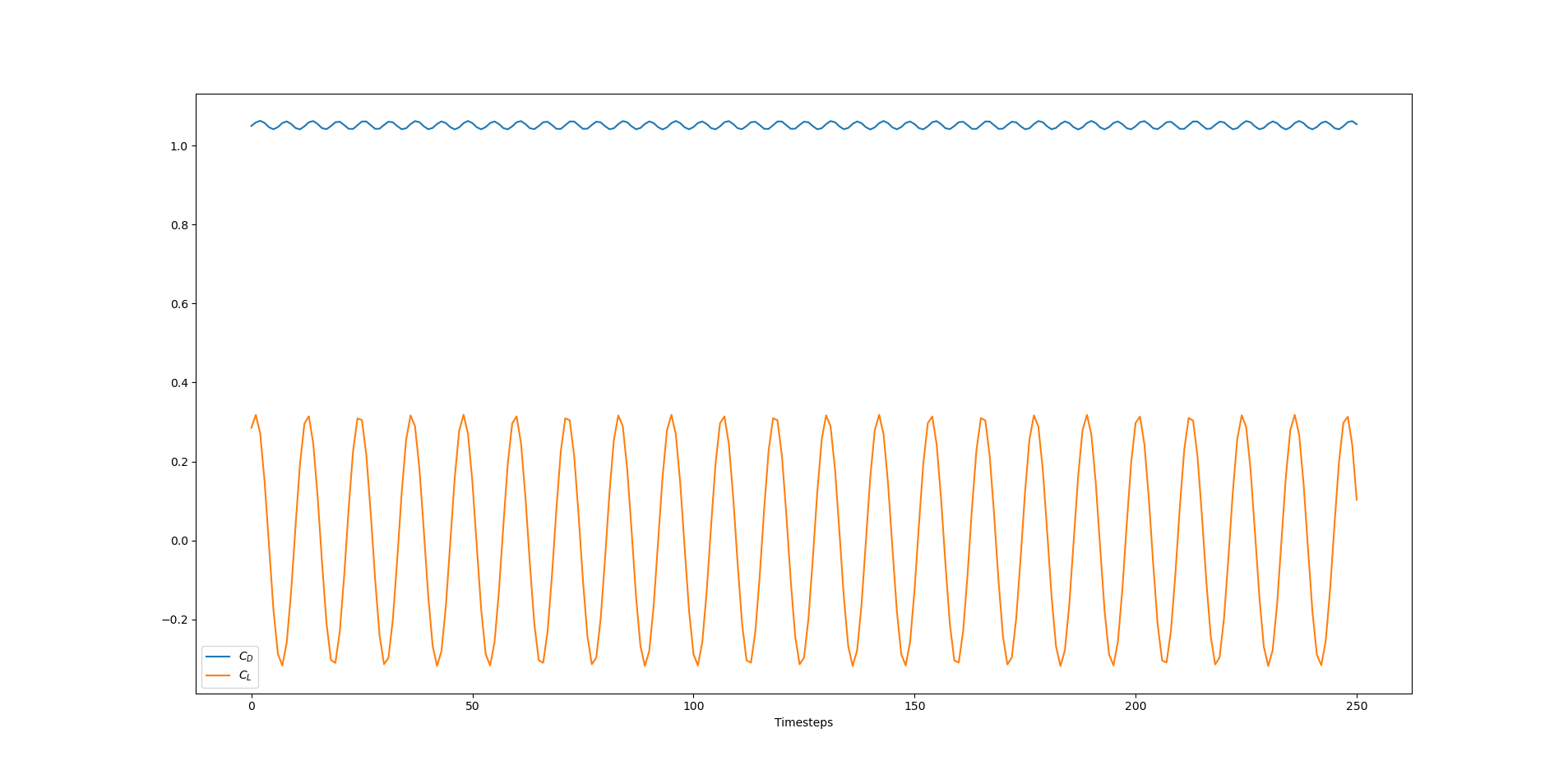}
            \label{fig:baseline_clcd}
        }
    }


    \caption{Uncontrolled flow at $\mathrm{Re} = 100$:
    $\bar{\widehat{C}}_D \approx 1.05$,
    $|\widehat{C}_L| \approx 0.31$, $St \approx 0.164$.}
    \label{fig:baseline}

\end{figure}

Fig.~\ref{fig:baseline} shows the uncontrolled cylinder wake at
$\mathrm{Re} = 100$. The flow settles onto a globally stable limit
cycle characterized by periodic Bénard-von Kármán vortex shedding,
with shedding frequency $f_s \approx 0.164$\,Hz (Strouhal number
$St = f_s D / U_\infty \approx 0.164$). The pressure-approximated
time-averaged force coefficients are $\bar{\widehat{C}}_D \approx 1.05$
and $\bar{\widehat{C}}_L \approx 0$ (by symmetry of the time average),
with peak lift amplitude $|\widehat{C}_L| \approx 0.31$. These values
serve as the baseline against which drag reduction is measured.


\begin{remark}\label{remark5}
    The globally stable limit cycle is a key structural feature that makes the rollout-based iLQR procedure particularly well-suited to this problem: when the controller is inactive between rollouts, the flow naturally returns to the vicinity of the shedding initial condition, eliminating the need for explicit state resetting between iLQR iterations. Thus, this approach can, in principle, be implemented online with a single rollout of the perturbed flow, with periodic resets to the vortex shedding flow.
\end{remark}

\subsection{Optimal Controls}
\label{subsec:results}

Fig.~\ref{fig:results} and Fig.~\ref{fig:vorticity} summarize the
controlled flow response across both stages.

\subsubsection*{Stage~I: POD-iLQR Transfer}
\begin{figure}[t]
    \subfloat[Episodic cost versus POD-iLQR iteration.]{
      \includegraphics[width=0.5\linewidth]{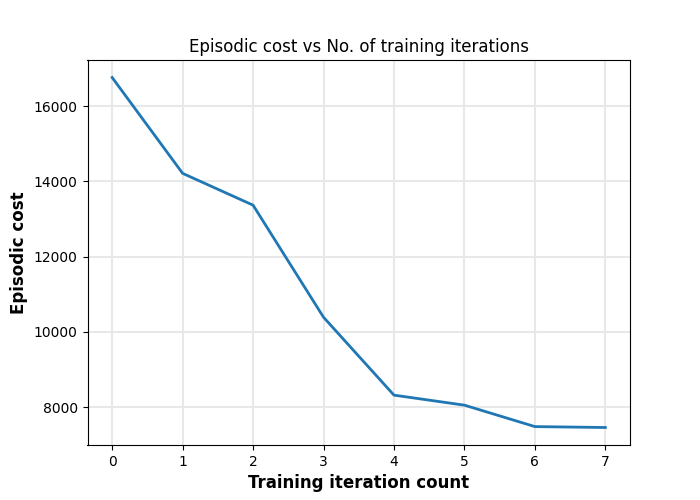}
      \label{fig:episodic_cost}
    }
    \subfloat[Optimal control sequence $u_k$.]{
      \includegraphics[width=0.5\linewidth]{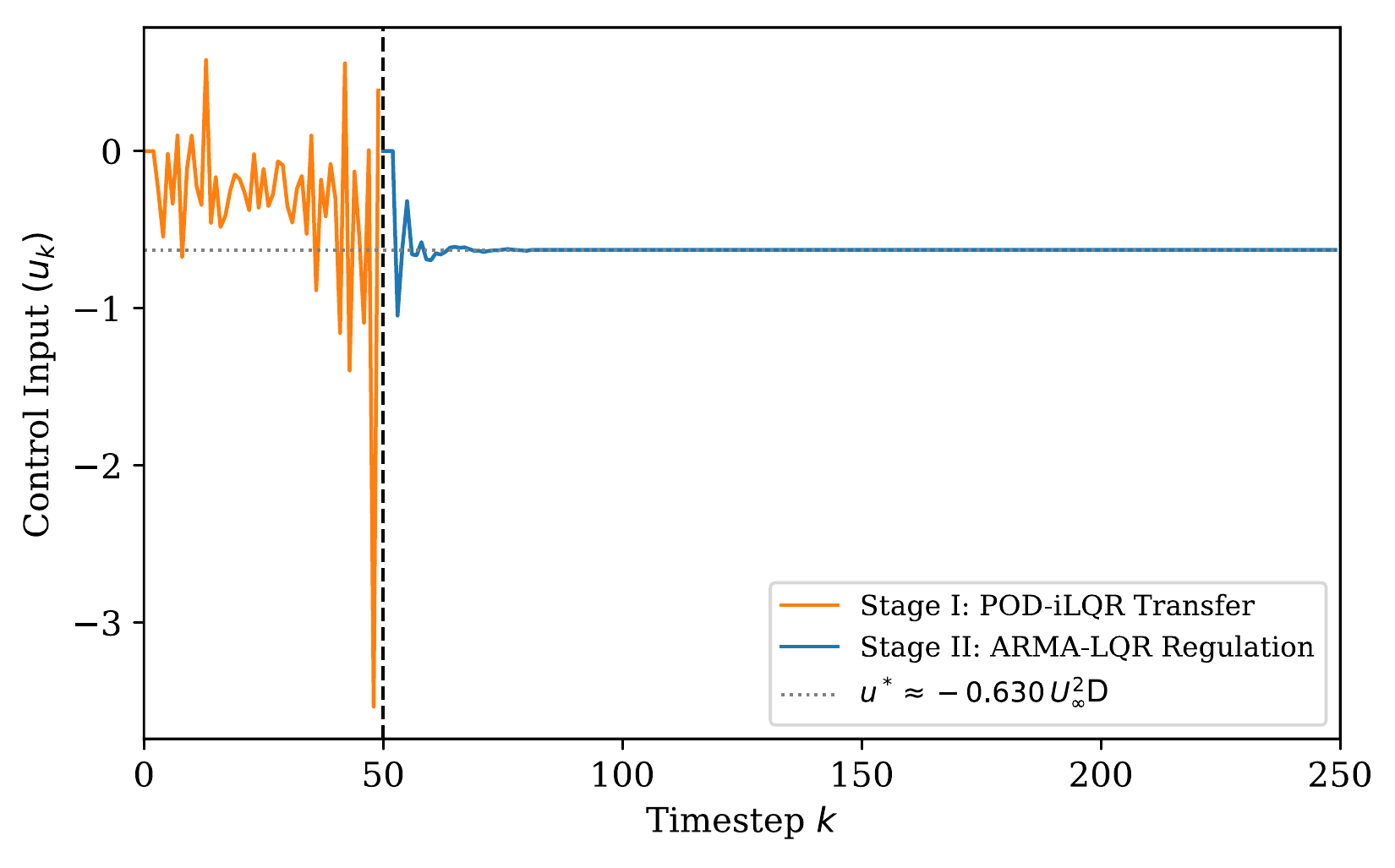}
      \label{fig:control_input}
    }
    \caption{(a) The episodic cost decreases monotonically over 7 iLQR
    iterations. (b) Optimal control sequence $u_k$ over the full
    two-stage trajectory. \textit{Orange (Stage~I):} time-varying
    POD-iLQR drives the flow off the limit cycle with aggressive,
    sign-alternating inputs over $N=50$ steps. \textit{Blue (Stage~II):}
    ARMA-LQR feedback converges to a near-constant output
    $u^*= -0.63 \,U_\infty^2 D$, representing the persistent
    antisymmetric blowing/suction required to maintain the operating point.}
    \label{fig:training_and_control}
\end{figure}

Starting from the limit-cycle initial condition, the POD-iLQR algorithm
is run for $7$ iterations. Fig.~\ref{fig:episodic_cost} shows the
episodic cost decreasing monotonically, with diminishing returns as
the trajectory approaches the equilibrium neighborhood; the algorithm
is terminated at a fixed iteration count. The optimal control sequence
at the final iteration is applied to PyFR. As shown in
Fig.~\ref{fig:results_cl}, $\widehat{C}_L$ is suppressed from peak
amplitude $|\widehat{C}_L| \approx 0.31$ to near zero over the
$N = 50$-step horizon. Simultaneously, $\widehat{C}_D$ decreases from
its limit-cycle mean $\bar{\widehat{C}}_D \approx 1.05$
(Fig.~\ref{fig:results_cd}), while the true drag coefficient
(Fig.~\ref{fig:results_cd_true}) drops from $\bar{C}_D \approx 1.42$,
confirming that the pressure-only surrogate tracks the full
surface-integrated force. The velocity snapshots
(Fig.~\ref{fig:vorticity}(a)-(c)) show the progressive collapse of the alternating vortex street over this horizon.

\subsubsection*{Stage~II: ARMA-LQR Stabilization}
\begin{figure*}[t]
    \subfloat[Pressure-approximated drag $\widehat{C}_D$.]
        {\includegraphics[width=0.25\textwidth]{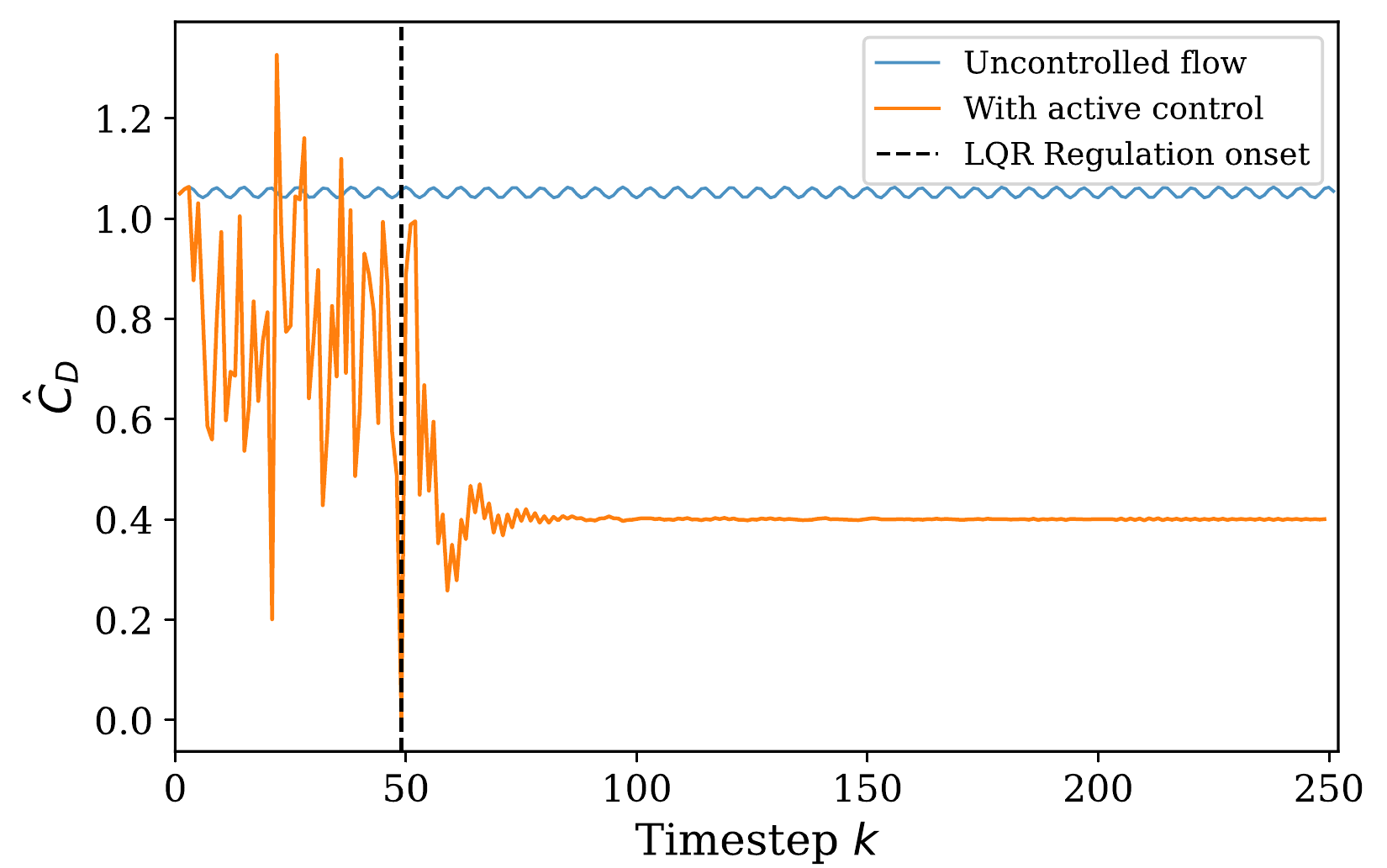}
        \label{fig:results_cd}}
    \subfloat[Pressure-approximated lift $\widehat{C}_L$.]
        {\includegraphics[width=0.25\textwidth]{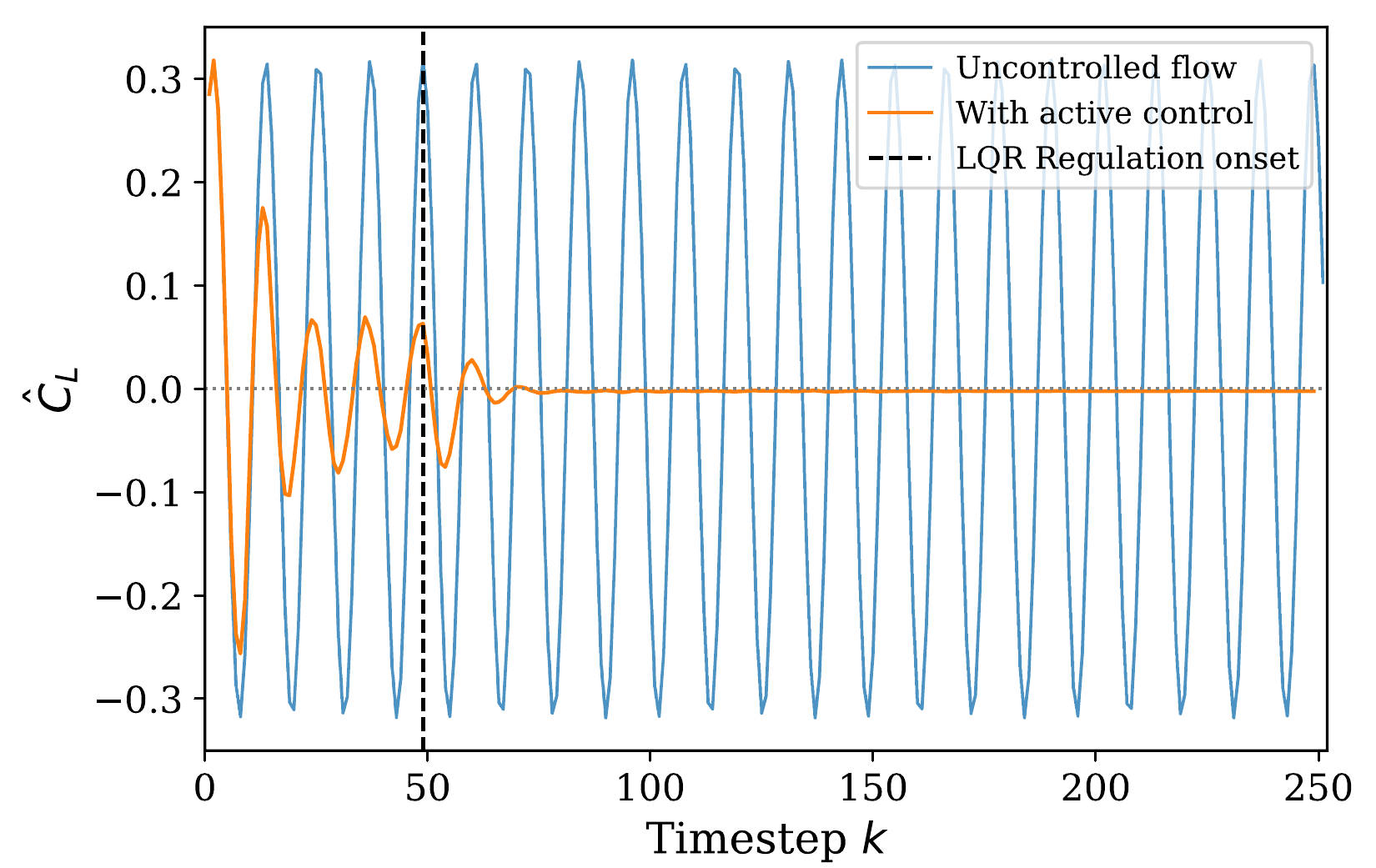}
        \label{fig:results_cl}}
    \subfloat[True drag $C_D$ (pressure + viscous).]
        {\includegraphics[width=0.25\textwidth]{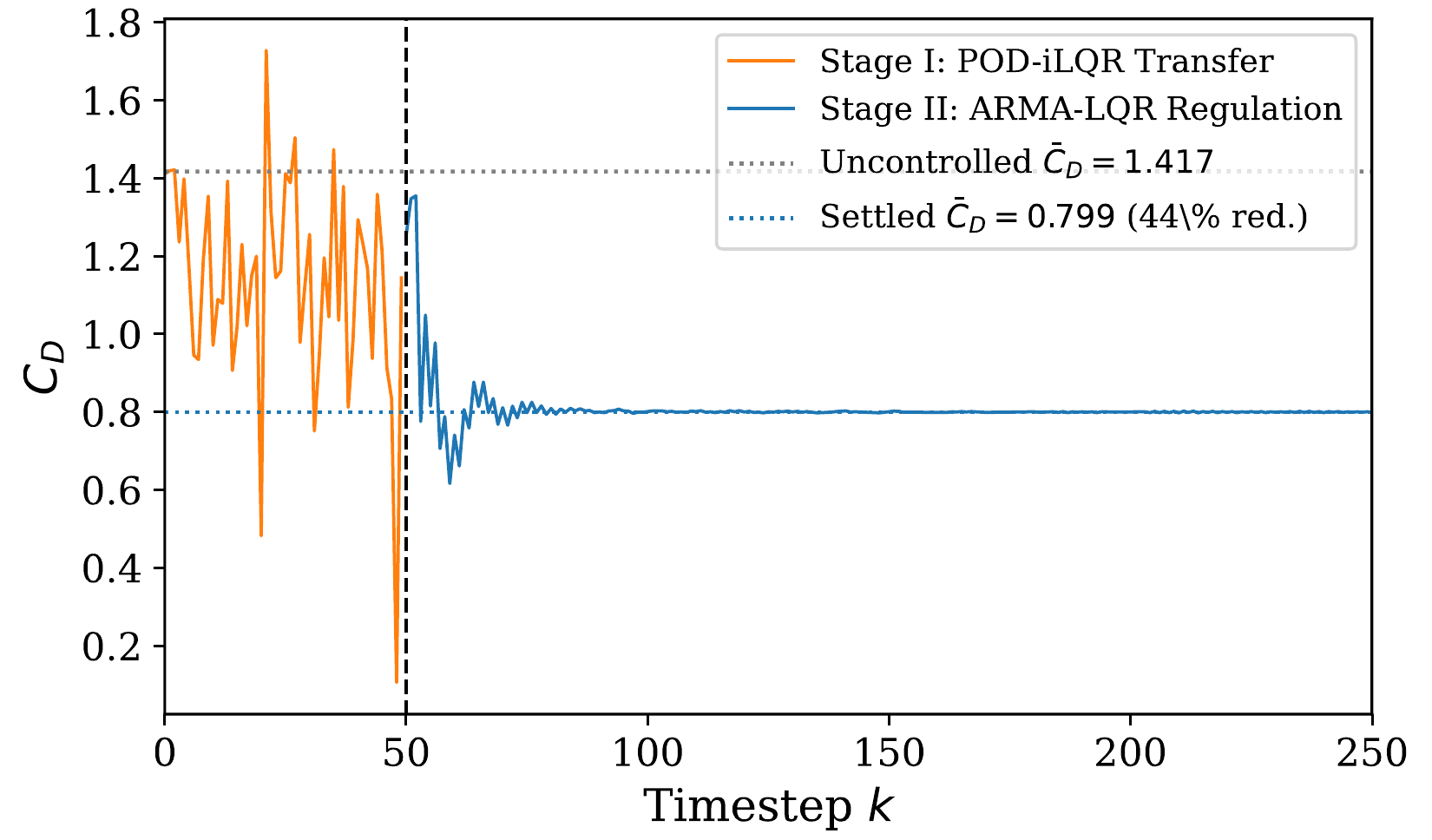}
        \label{fig:results_cd_true}}
    \subfloat[True lift $C_L$ (pressure + viscous).]
        {\includegraphics[width=0.25\textwidth]{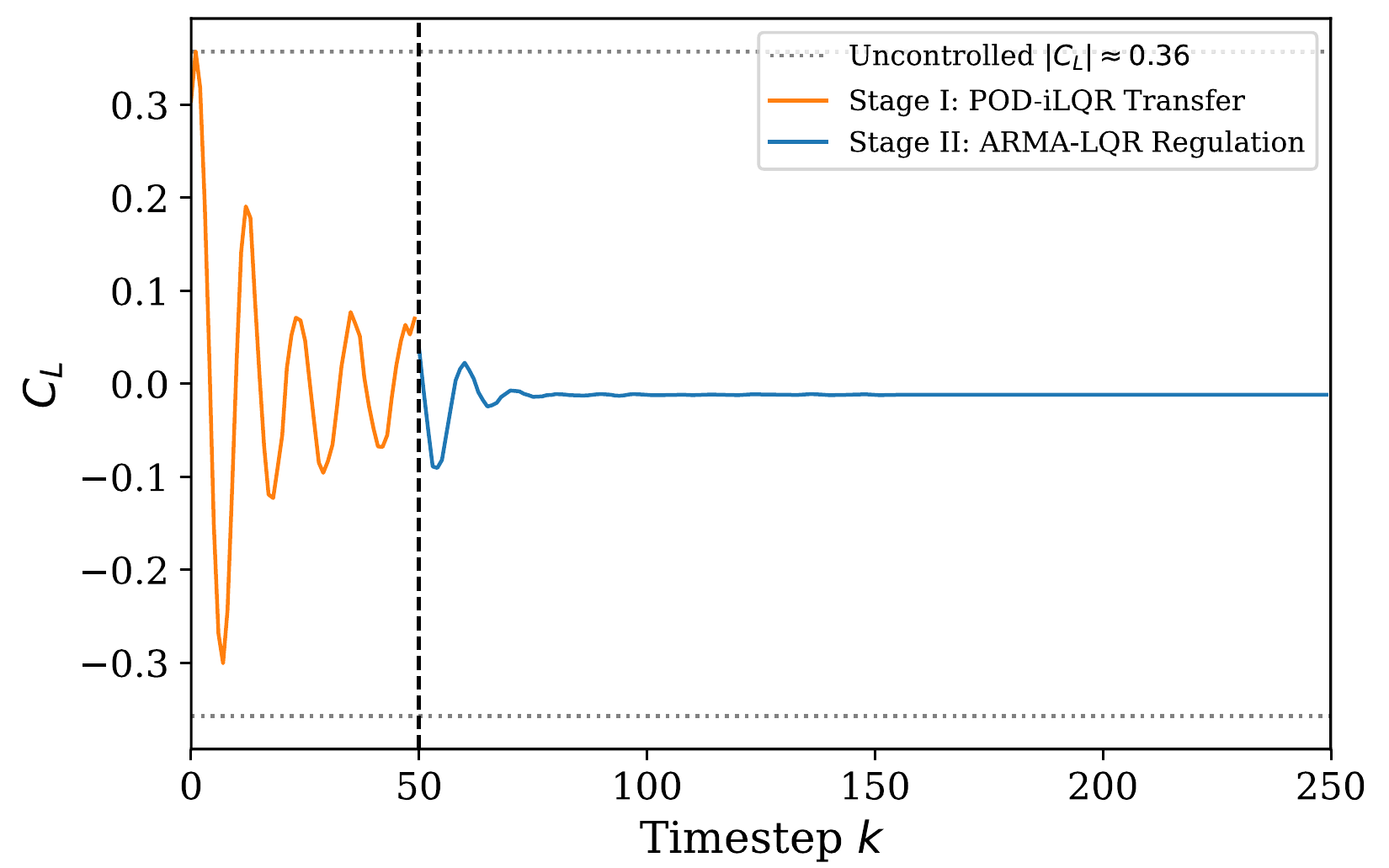}
        \label{fig:results_cl_true}}
    \caption{Force coefficients under the two-stage controller.
    Blue: uncontrolled baseline; orange: controlled response. Dashed vertical line at step~50 marks the Stage~I\,/\,Stage~II handoff.
    \textbf{(a)}~$\widehat{C}_D$ drops from $\bar{\widehat{C}}_D \approx 1.05$
    and stabilizes near $\widehat{C}_D^* \approx 0.40$ under Stage~II.
    \textbf{(b)}~$|\widehat{C}_L|$ is suppressed from $\approx 0.31$ to near zero within the Stage~I horizon and maintained through Stage~II.
    \textbf{(c)}~True drag (full surface integration) settles to $C_D^* = 0.799$, a $44\%$ reduction from the uncontrolled baseline
    $\bar{C}_D = 1.417$. The gap between $\widehat{C}_D^*$ and $C_D^*$
    reflects the omitted viscous contribution
    in Eqs.~\eqref{eq:cdhat}-\eqref{eq:clhat}.}
    \label{fig:results}
\end{figure*}

\begin{figure*}[t]
    \centering
    \subfloat[Stage~I, initial]
        {\includegraphics[width=0.3\textwidth]{figures/ilqr_norm1.png}
        \label{fig:vort_ilqr1}}
    \subfloat[Stage~I, mid-transfer]
        {\includegraphics[width=0.3\textwidth]{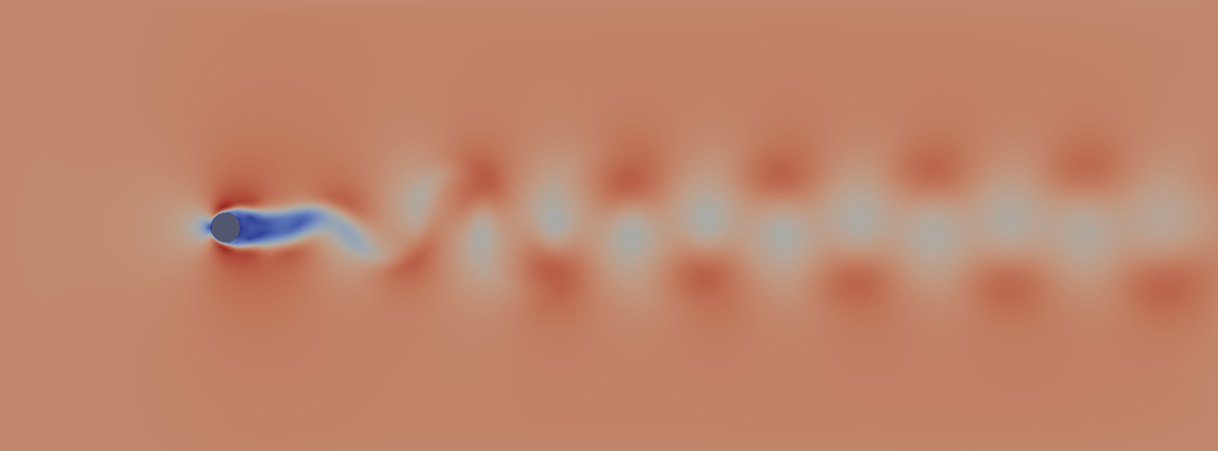}
        \label{fig:vort_ilqr2}}
    \subfloat[Stage~I, terminal]
        {\includegraphics[width=0.3\textwidth]{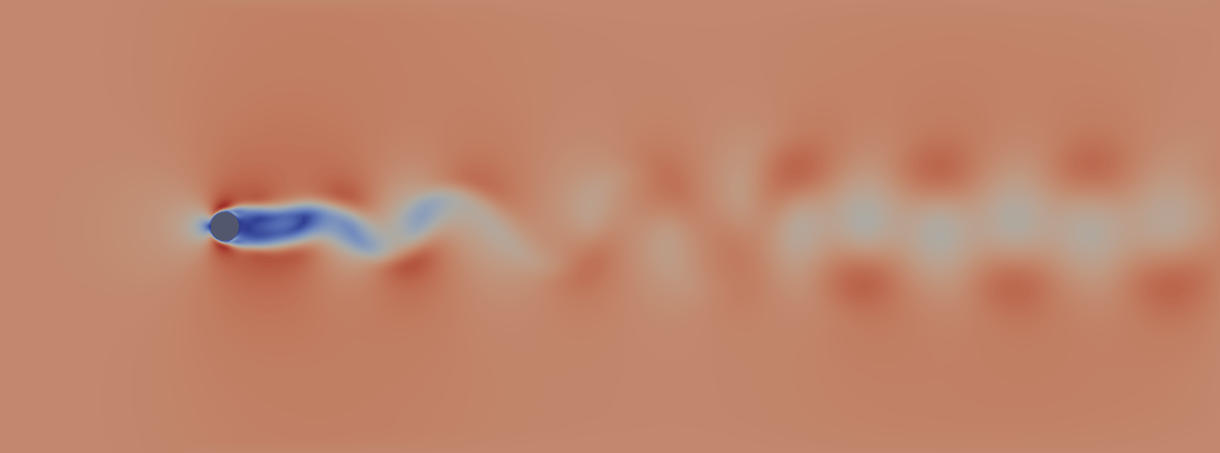}
        \label{fig:vort_ilqr3}}
    \\
    \subfloat[Stage~II, early regulation]
        {\includegraphics[width=0.3\textwidth]{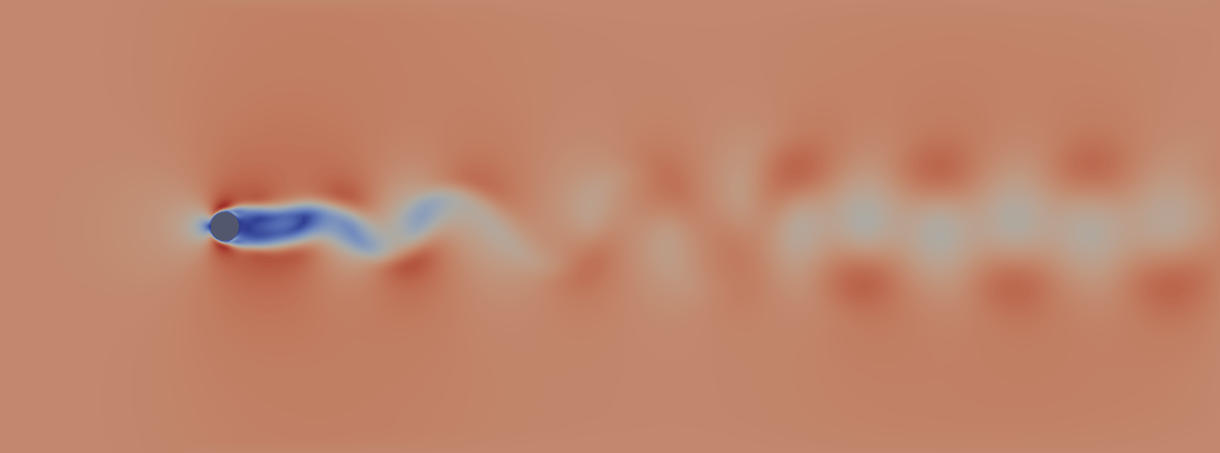}
        \label{fig:vort_lqr1}}
    \subfloat[Stage~II, intermediate]
        {\includegraphics[width=0.3\textwidth]{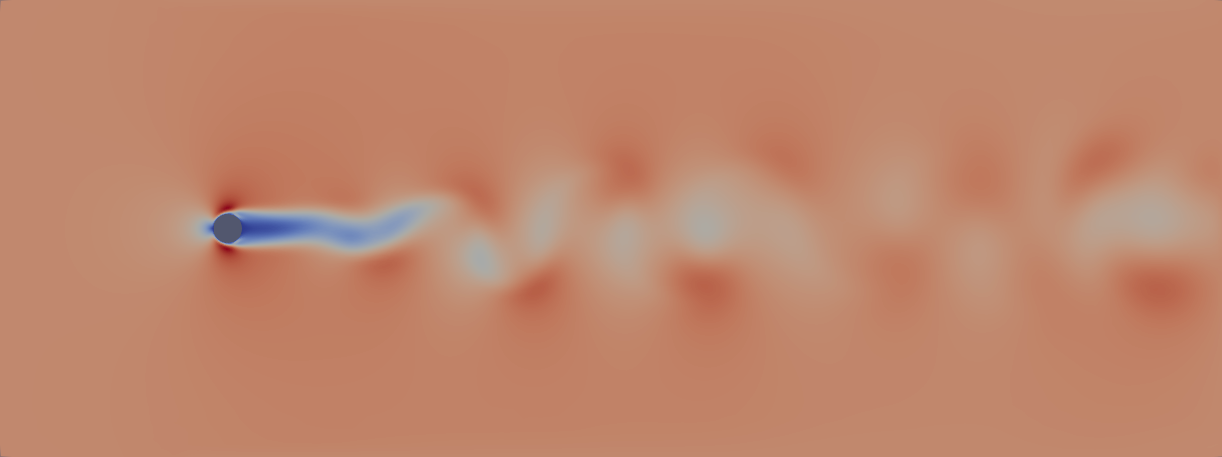}
        \label{fig:vort_lqr2}}
    \subfloat[Stage~II, steady state]
        {\includegraphics[width=0.3\textwidth]{figures/lqr_norm3.png}
        \label{fig:vort_lqrend}}
    \\
    \subfloat[No regulation, initial]
        {\includegraphics[width=0.3\textwidth]{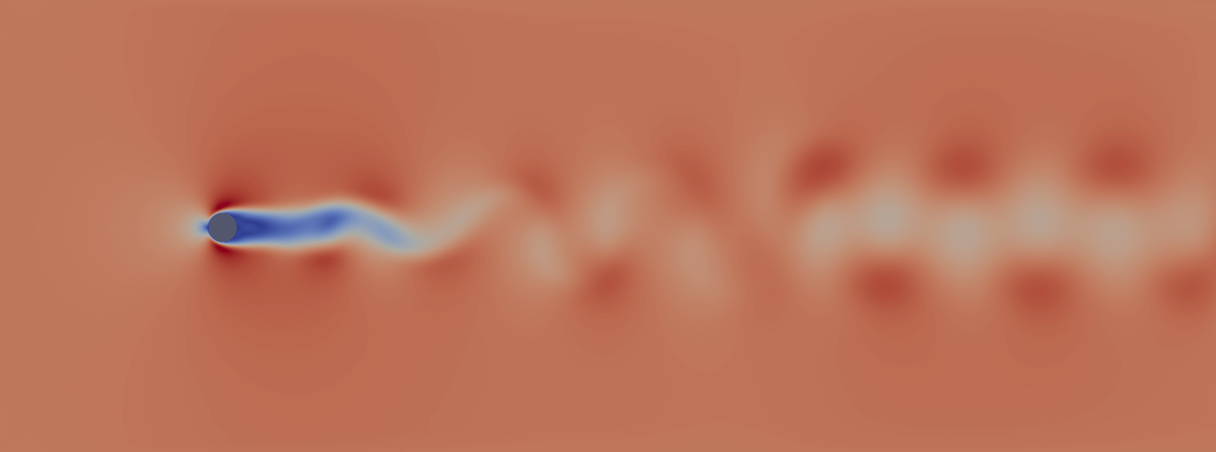}
        \label{fig:no_reg1}}
    \subfloat[No regulation, intermediate]
        {\includegraphics[width=0.3\textwidth]{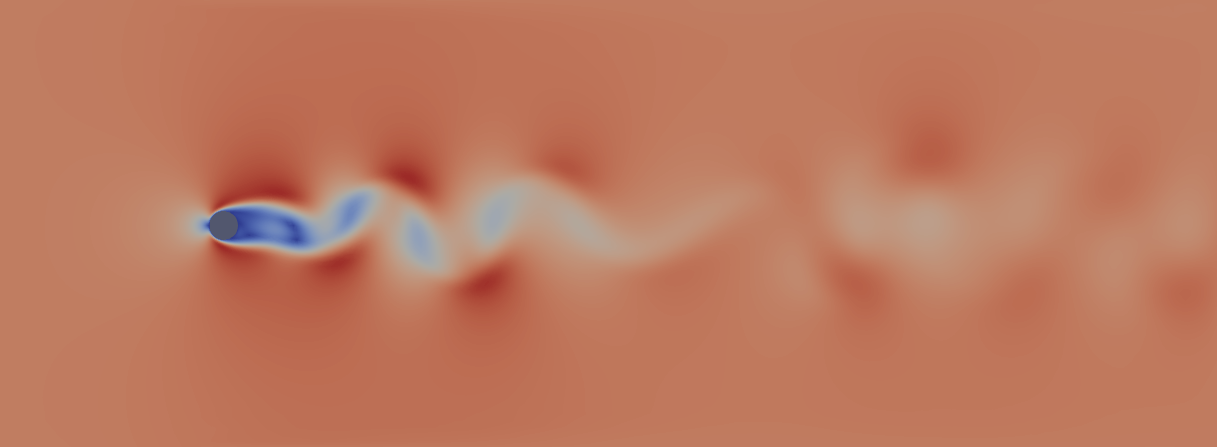}
        \label{fig:no_reg2}}
    \subfloat[No regulation, limit cycle]
        {\includegraphics[width=0.3\textwidth]{figures/uncontrolled_flow_updated.png}
        \label{fig:no_reg3}}
    \\
    \subfloat[ARMA-LQR only, initial]
        {\includegraphics[width=0.3\textwidth]{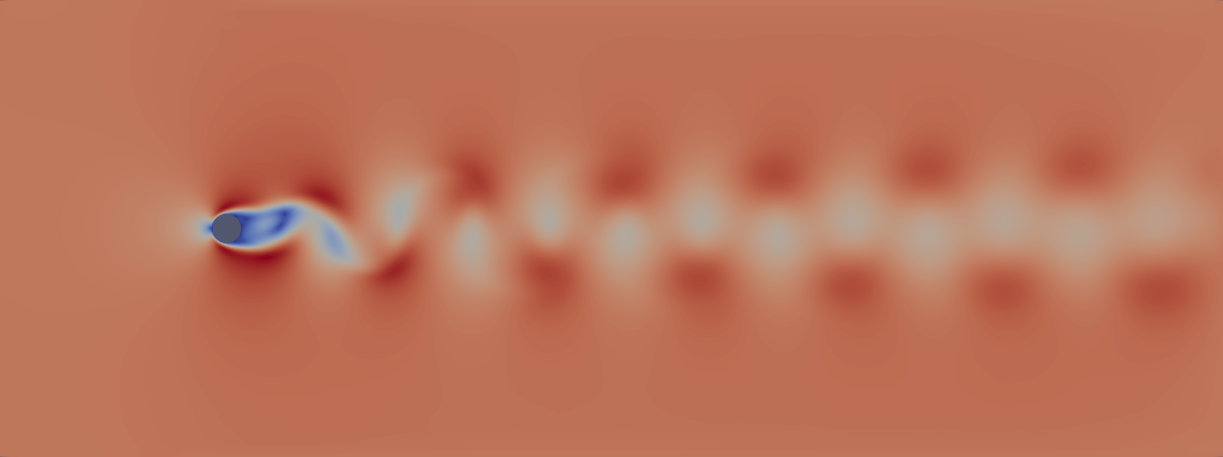}
        \label{fig:lqr_only1}}
    \subfloat[ARMA-LQR only, intermediate]
        {\includegraphics[width=0.3\textwidth]{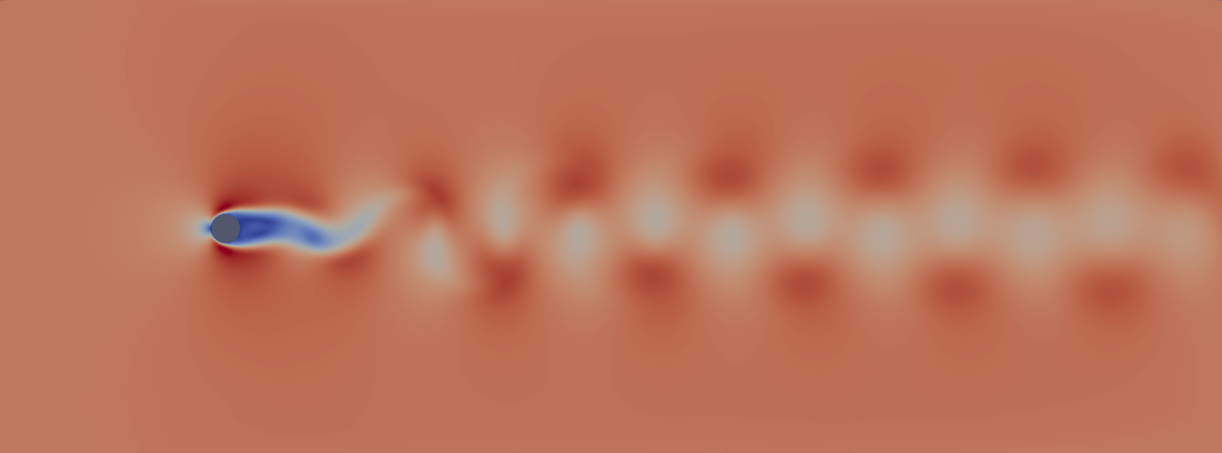}
        \label{fig:lqr_only2}}
    \subfloat[ARMA-LQR only, terminal]
        {\includegraphics[width=0.3\textwidth]{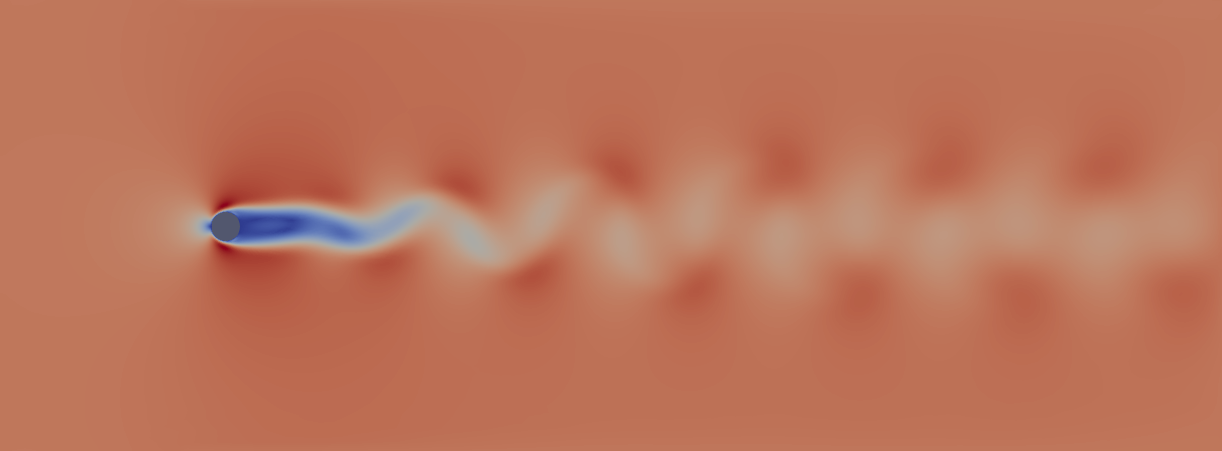}
        \label{fig:lqr_only3}}
    \caption{Instantaneous streamwise velocity field showing progressive
    suppression of the Bénard-von Kármán vortex street.
    \textbf{(a)-(c):} Stage~I POD-iLQR transfer drives the flow from
    sustained shedding to the equilibrium neighborhood over 50 steps.
    \textbf{(d)-(f):} Stage~II ARMA-LQR regulation stabilizes the flow
    to the symmetric steady wake.
    \textbf{(g)-(i):} If there is no regulation, the unstable equilibrium quickly falls back into the nonlinear limit cycle (vortex shedding).
    \textbf{(j)-(l):} ARMA-LQR applied without Stage~I transfer:
    the controller reduces shedding amplitude but cannot drive the flow to the symmetric equilibrium, as the linearization operates outside the equilibrium's basin of attraction. Residual vortex shedding
    persists throughout, confirming that Stage~I is necessary for full suppression.}
    \label{fig:vorticity}
\end{figure*}

Beginning from the terminal Stage~I snapshot, the ARMA-LTI model is identified from $n_r = 10$ rollouts of $H = 50$ steps. The LQR regulator drives both $\widehat{C}_D$ and $\widehat{C}_L$ to steady values within approximately 50 additional steps (Fig.~\ref{fig:results_cd}, \ref{fig:results_cl}). The Stage~II controller settles to a near-constant output $u^* \approx -0.63\,U_\infty^2 D$, representing the persistent forcing required to maintain the flow at the unstable equilibrium against its natural tendency to return to the shedding limit cycle~\cite{Illingworth2016}. The pressure-approximated drag stabilizes near $\widehat{C}_D^* \approx 0.40$, while the true drag coefficient, computed from full surface integration of pressure and viscous contributions, settles to $C_D^* = 0.799$
(Fig.~\ref{fig:results} (c)), a $44\%$ reduction relative to the uncontrolled baseline $\bar{C}_D = 1.417$. The discrepancy between $\widehat{C}_D^*$ and $C_D^*$ reflects the omitted viscous contribution in Eqs.~\eqref{eq:cdhat}-\eqref{eq:clhat}; at $\mathrm{Re}=100$,
pressure drag dominates the total drag, so the surrogate provides a consistent feedback signal despite underestimating the absolute level.
Fig.~\ref{fig:vorticity} (d)-(f) confirm complete suppression of the vortex street, with the steady symmetric wake characteristic of the unstable equilibrium. 


\begin{figure*}[t]
    \centering
    \subfloat[ARMA-LQR only: $\widehat{C}_D$.]
        {\includegraphics[width=0.27\textwidth]{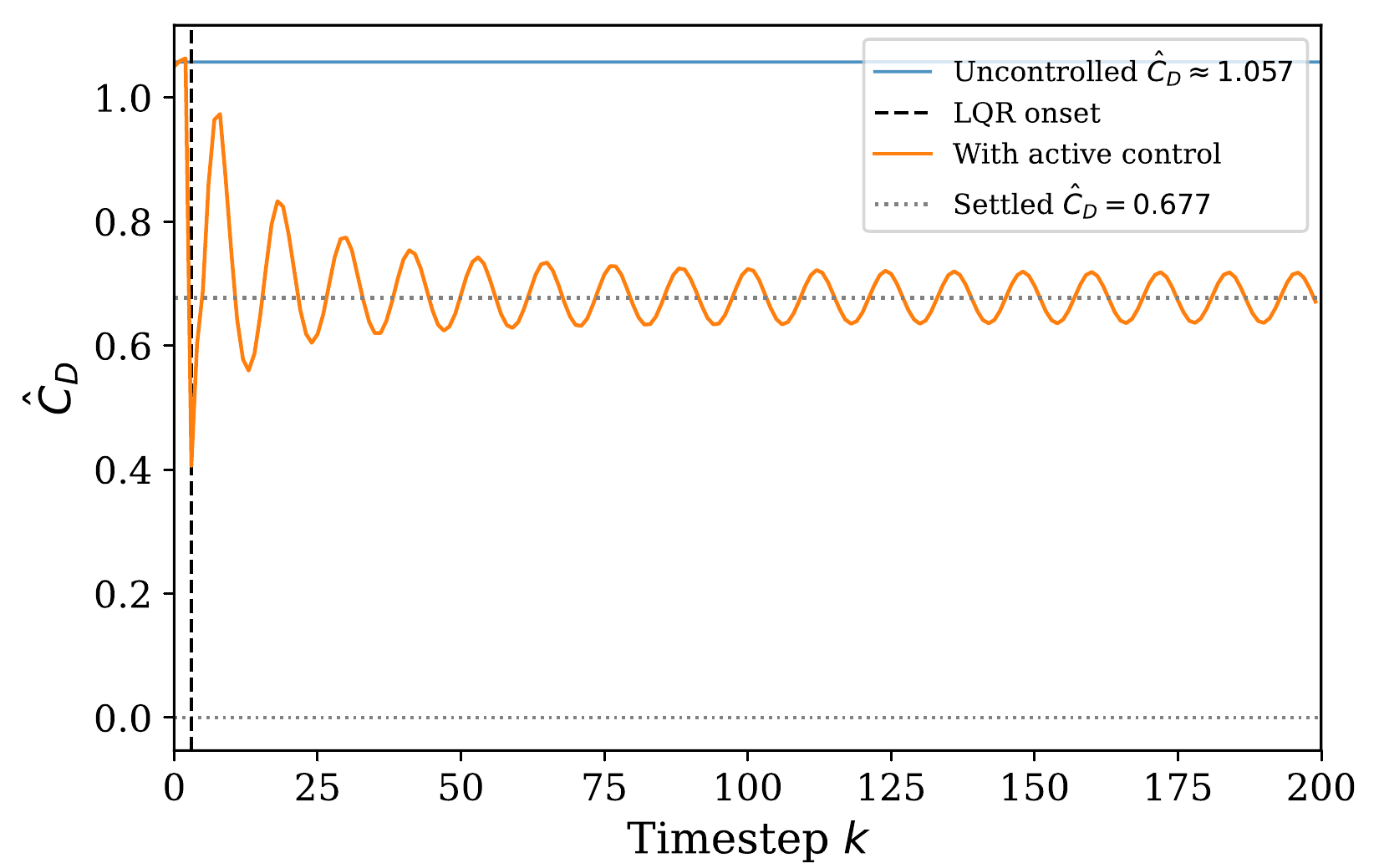}
        \label{fig:lqr_only_cdhat}}
    \subfloat[ARMA-LQR only: $\widehat{C}_L$.]
        {\includegraphics[width=0.27\textwidth]{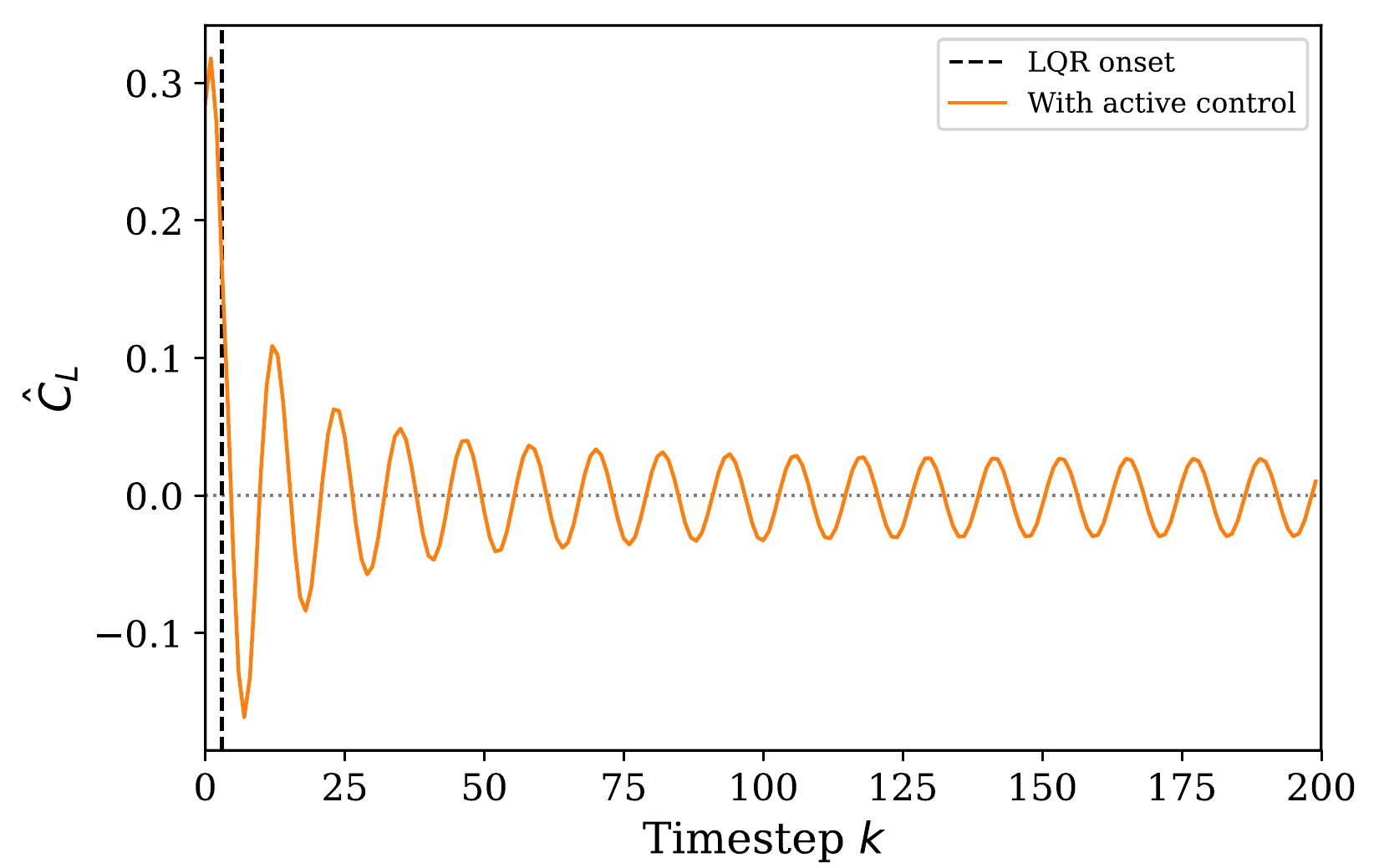}
        \label{fig:lqr_only_clhat}}
    \subfloat[ARMA-LQR only: true $C_D$.]
        {\includegraphics[width=0.27\textwidth]{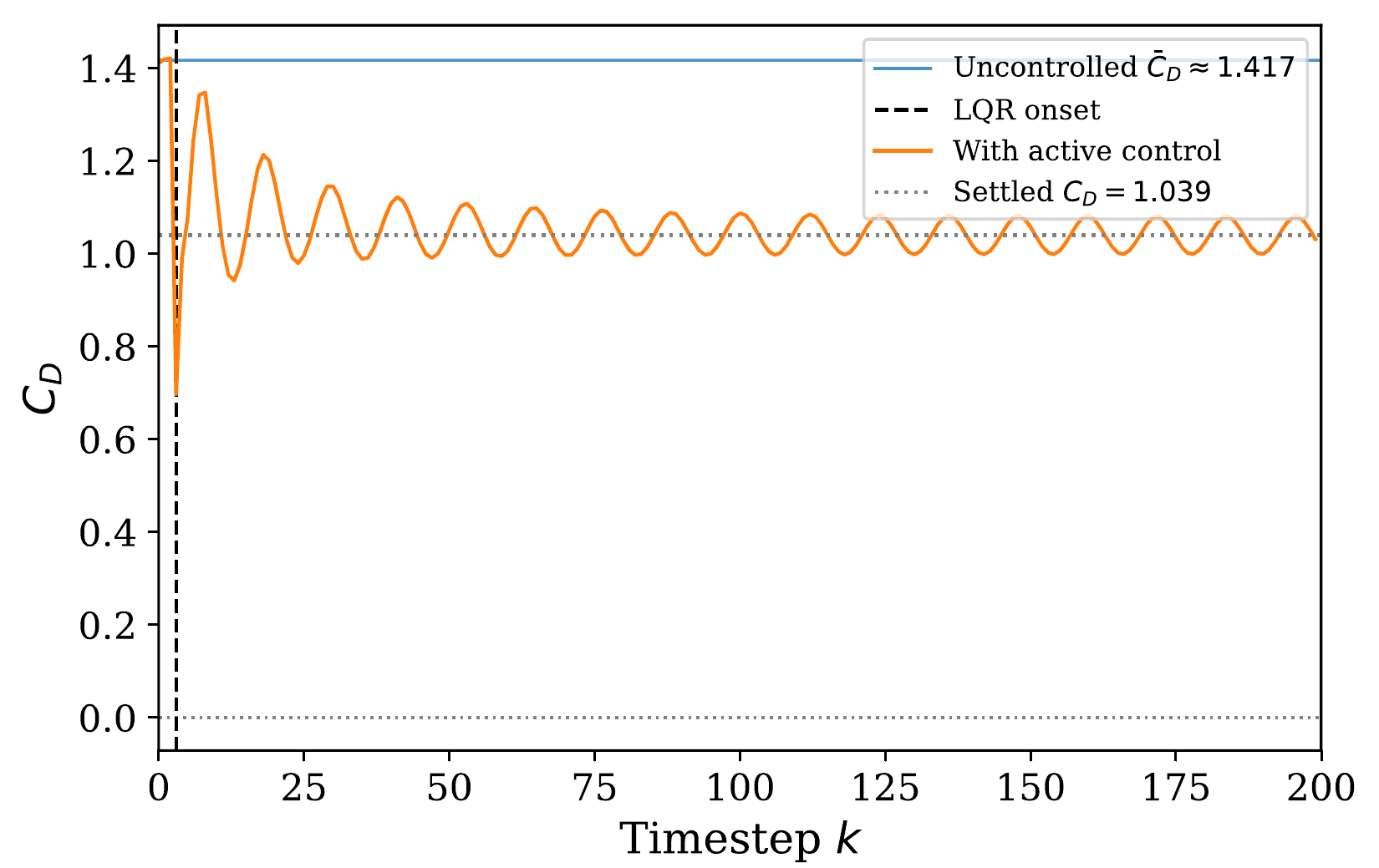}
        \label{fig:lqr_only_cd}}
    \caption{Force coefficients under ARMA-LQR regulation alone,
    without Stage~I POD-iLQR transfer (cf.\ Fig.~\ref{fig:vorticity} (j)-(l)).
    \textbf{(a)}~$\widehat{C}_D$ decreases from the limit-cycle mean but does not converge to the equilibrium value.
    \textbf{(b)}~$\widehat{C}_L$ is partially attenuated yet retains persistent residual oscillations, confirming that the flow remains on a perturbed limit cycle rather than at the symmetric equilibrium.
    \textbf{(c)}~True drag settles to $C_D \approx 1.04$, a $\sim\!27\%$ reduction compared to $44\%$ achieved by the full two-stage scheme,
    demonstrating the necessity of Stage~I for complete suppression.}
    \label{fig:results_lqronly}
\end{figure*}

\subsection{Discussion}
\label{subsec:discussion}
The results demonstrate complete suppression of vortex shedding with a $44\%$ reduction in total drag ($C_D^*: 1.417 \to 0.799$, pressure +
viscous, full surface integration; Fig.~\ref{fig:results}(c))
relative to the uncontrolled baseline, using only eight surface pressure sensors and no adjoint solver, reduced-order model, or full-state access~\cite{TNNLS_PODiLQR}. The pressure-approximated surrogate used for feedback shows a consistent reduction ($\widehat{C}_D^*: 1.05 \to 0.4$; Fig.~\ref{fig:results}), with the gap between $C_D^*$ and $\widehat{C}_D^*$ attributable to the omitted viscous contribution in Eqs.~\eqref{eq:cdhat}-\eqref{eq:clhat}. At $\mathrm{Re}=100$, pressure drag dominates the total drag for a circular cylinder \cite{Batchelor1967}
, so the surrogate provides a reliable feedback signal; at higher Reynolds numbers, where viscous effects become relatively smaller, the approximation improves further. 

The two-stage decomposition is essential to this result. The POD-iLQR stage handles the global nonlinear transfer: without it, a linear regulator initialized from the shedding limit cycle would fail to navigate the nonlinear landscape separating the limit cycle from the unstable equilibrium, as demonstrated quantitatively in 
Fig.~\ref{fig:results_lqronly}, where ARMA-LQR alone achieves only ${\sim}27\%$ drag reduction with persistent residual shedding. Conversely, the ARMA-LQR stage is necessary for asymptotic stabilization: the finite-horizon POD-iLQR trajectory brings the flow into the neighborhood of the equilibrium but cannot guarantee regulation, as evidenced by the ``no regulation'' evolution in Fig.~\ref{fig:vorticity} (g)-(i), where the flow returns to the shedding limit cycle when feedback is removed.


In Stage~II, we regulate to a zero reference in the surrogate force coordinates, i.e., $z^*=[0,0]^\top$ for $(\widehat{C}_D,\widehat{C}_L)$. Here $\widehat{C}_L=0$ encodes the steady symmetric wake, while $\widehat{C}_D$ is included only as a soft performance term. Accordingly, the terminal value $\widehat{C}_{D,N}$ reached at the end of Stage~I is not a prescribed setpoint; it simply indicates the operating point delivered by the transfer and serves as the linearization/identification point for Stage~II. The regulator then drives the system to its locally stabilized equilibrium, yielding the final $\widehat{C}_D^*$. We therefore weight lift much more heavily than drag ($w_{\widehat{C}_L}\gg w_{\widehat{C}_D}$, with stronger emphasis in Stage~II): $\widehat{C}_L=0$ is the reliable equilibrium signature, whereas drag reduction is an outcome of stabilization under limited actuation rather than a hard target.

\begin{remark}
    Due to paucity of space, we omit the additional validation at a higher Reynolds number (Re = 150) from the main text; complete results are provided in our public repository (including scripts, parameters, and representative force histories/flow visualizations), which reproduce the same qualitative two-stage behavior and support reproducibility beyond the Re = 100 case reported here. The two-stage architecture results in a complete lift suppression and a total drag reduction of 49.2\%. 
\end{remark}

\section{Conclusions}
We presented a data-driven, output-feedback framework for suppressing
vortex shedding in the cylinder wake at $\mathrm{Re}=100$ using only
eight surface pressure sensors. The two-stage architecture achieves
complete lift suppression and a $44\%$ total drag reduction without an adjoint solver, reduced-order model, or full-state access. The decomposition is
essential: Stage~I navigates the nonlinear landscape that a local
regulator cannot traverse, while Stage~II provides the asymptotic
stabilization that a finite-horizon optimizer cannot guarantee.

The limit cycle recurrence property identified in
Remark~\ref{remark5} has a direct implication for scalability: because
deactivating the controller resets the flow to a known distributional
initial condition, the identification-control loop can be executed
online without a simulator. Future work will exploit this property in two directions: (i)~developing an online learning variant in which ARMA models are continuously refined across physical rollouts, eliminating the need for a simulator and enabling direct deployment
in wind- and water-tunnel experiments on airfoils, non-circular bluff bodies, and other practically relevant geometries; and (ii)~scaling to higher Reynolds numbers where the limit-cycle structure persists but richer dynamics demand online model adaptation.

\bibliographystyle{IEEEtran}
\bibliography{IEEEabrv,ICRA_refs,TAC_refs}



\end{document}